\documentclass[pre,twocolumn,showpacs,preprintnumbers]{revtex4-1}
\usepackage{amsmath,amssymb,latexsym} 
\usepackage{color}

\usepackage{graphicx} 
\usepackage{natbib}
\usepackage{epstopdf}
\usepackage{color}
\usepackage{makecell}
\usepackage[table]{xcolor}
\usepackage{tabularx}
\usepackage{ragged2e}

\begin{document}

\newcommand{\LL}[1]{\textcolor{red}{{\bf LL:} #1}}
\newcommand{\LH}[1]{\textcolor{blue}{{\bf NM:} #1}}
\newcommand{\CH}[1]{\textcolor{cyan}{{\bf Ch:} #1}}

\title
{Isochoric, isobaric and ultrafast conductivities of\\ aluminum, lithium
and carbon 
in the warm dense matter regime
}

\author
{M.W.C. Dharma-wardana\email[Email address:\ ]
{chandre.dharma-wardana@nrc-cnrc.gc.ca} and D. D. Klug}
\affiliation{
National Research Council of Canada, Ottawa, Canada, K1A 0R6
}

\author{L. Harbour and Laurent J. Lewis}
\affiliation{
D\'{e}partement de Physique, Universit\'{e}
 de Montr\'{e}al, Montr\'{e}al, Qu\'{e}bec, Canada.} 

%
\date{\today}

\begin{abstract}
We study the conductivities $\sigma$ of (i) the  equilibrium isochoric state
($\sigma_{\rm is}$),  (ii) the equilibrium isobaric state 
($\sigma_{\rm ib}$),  and also the
(iii) non-equilibrium ultrafast matter (UFM) state ($\sigma_{\rm uf}$) with
the ion temperature $T_i$ less than the the electron temperature $T_e$.
Aluminum, lithium  and carbon are considered, being
increasingly complex warm dense matter (WDM)
systems, with carbon having transient covalent bonds. First-principles calculations, i.e.,
neutral-pseudoatom (NPA) calculations and  density-functional theory (DFT) with
molecular-dynamics (MD) simulations, are compared where possible with experimental data to
characterize $\sigma_{\rm ic}, \sigma_{\rm ib}$ and $\sigma_{\rm uf}$. The NPA
$\sigma_{\rm ib}$ are closest to the  available experimental data 
when compared to results from DFT+MD, 
where simulations of about 64-125 atoms are typically used.  
 The published conductivities
for  Li are reviewed and the value at a temperature of 4.5 eV 
is examined using  supporting X-ray
Thomson scattering calculations. A physical
picture of the variations of $\sigma$ with temperature and density applicable to
these materials is given.
The insensitivity of $\sigma$ to $T_e$ below 10 eV for carbon,  compared to Al and
Li, is clarified. 
\end{abstract}

\maketitle
\section{Introduction}
\label{intro} 
Short-pulsed lasers  as well as shock-wave  techniques can probe matter in
hitherto experimentally inaccessible regimes of great interest. These provide
information needed for understanding normal matter  and unusual states of
matter, in equilibrium or in transient conditions~\cite{Ng2011, Milchberg88}.
Similar `hot-carrier' processes occur in semiconductor
nanostructures~\cite{DharmaHot93, Vaziri2013}.  Such warm dense matter (WDM)
systems include not only equilibrium  systems where the ion temperature $T_i$ and
the electron temperature $T_e$ are equal, but also systems where $T_i \ne T_e$,
or highly non-equilibrium systems where the notion of temperature is
inapplicable~\cite{Medvedev}. While the prediction of a  quasi equation of
state (quasi EOS) and related static properties for two-temperature (2$T$)
systems~\cite{HarbPhon17} is satisfactory, the conductivity calculations using
standard codes, even for sodium at the melting point,   require  massive quantum
simulations with as much as $\sim$1500 atoms and over 56 $k$-points (according to
Ref.~\cite{Pozzo11}), 
whereas even theories of the 1980s evaluated the sodium conductivities successfully
via a momentum relaxation-time ($\tau_{\rm mr}$) approach~\cite{Sinha89}, which 
is also used in Drude fits to the Kubo-Greenwood (KG) formula  used
with density-functional theory (DFT) and  molecular dynamics (MD) methods. The
KG-formula and its scope  are discussed further in the Appendix.


The static electrical conductivities of WDM equilibrium systems (i.e., $T_i=T_e$), as
well as 2$T$ quasiequilibrium systems, are the object of the present study. We
distinguish the isobaric equilibrium conductivity $\sigma_{\rm ib}$ and the
isochoric equilibrium conductivity $\sigma_{\rm ic}$  from the ultrafast matter
(UFM) quasiequilibrium (isochoric) conductivity $\sigma_{\rm uf}$. The 2$T$ WDM
states exist only for times  shorter than the electron-ion equilibration time
$\tau_{\rm ei}$ and may be accessed using femtosecond probes. 

We consider three systems of increasing complexity above the melting point: 
(a) a `simple' system, viz., WDM-aluminum at density $\rho$ =2.7 g/cm$^3$; (b)
WDM-lithium at 0.542 g/cm$^3$; and (c) WDM-carbon (2.0-3.7 g/cm$^3$) including
the low-$T$ covalent-bonding regime. As experimental data are available for
the isobaric evolution of Al and Li starting from their nominal normal densities
and down to lower densities of the expanded fluid, we
calculate $\sigma_{\rm ib}$ for Al and Li. The ultrafast conductivity
 $\sigma_{\rm uf}$
is calculated for all three materials, as $\sigma_{\rm is}$ is conveniently
accessible via short-pulse laser experiments.

The electrons in WDM-Li  are known to be
non-local with complex interaction effects. For instance, clustering effects
may appear~\cite{bonev2008} as the density is increased. WDM-carbon is a
complex liquid with transient covalent bonding where the C-C bond energy
$E_{cc}$ may reach $\sim 8$ eV in dilute gases. The three  conductivities 
$\sigma_{\rm ic}$,  $\sigma_{\rm ib}$, and  $\sigma_{\rm uf}$ for Al, Li and
C,  are calculated via two first-principles methods where, however, both finally
use a `mean-free path' model to estimate the conductivity. The two methods are:
(i) the neutral
pseudoatom (NPA) method as formulated by Perrot and Dharma-wardana
~\cite{HarbPhon17, CPP,eos95,Pe-Be} together with the Ziman  formula, and  (ii)
the  DFT+MD and KG  approach as available in codes such as VASP and
ABINIT~\cite{codes}, enabling us to assess the  extent of the agreement among
these theoretical methods and the available experiments. The liquid-metal experimental
data are still the most accurate data on WDM systems
available; they are used where possible to compare with
 calculations.

Accurate experimental data for the isobaric liquid state of
 Al~\cite{IJTher83,Gathers86}
and Li~\cite{ORNL88} are available, and provide a test of the theory. No
reliable isobaric carbon data are available; carbon at 3.7-3.9 g/cm$^3$ and
100-175 GPa was studied recently by x-ray Thomson scattering (XRTS)
~\cite{kraus13}. Hence we evaluate only $\sigma_{\rm ic}$ and $\sigma_{\rm
ib}$ in this case, for $\rho$ in the range of XRTS experiments and  related
simulations~\cite{Whitley15}. The conductivity  across a recently-proposed
phase transition~\cite{cdw-carbon16} in low-density carbon ($\sim 1.0$
g/cm$^3$) near $T\simeq 7$ eV is not addressed here.      

DFT+MD methods treat hot plasmas as a thermally evolved sequence of frozen
solids with a periodic unit cell of $N$ atoms --- typically $N\sim 100$,
although  order-of-magnitude larger systems may be needed~\cite{Pozzo11} for
reliable  transport calculations. The static conductivity $\sigma$  is
 evaluated from the
$\omega\to0$ limit of the KG $\sigma(\omega)$  using a  phenomenological model
(e.g.,  the Drude $\sigma(\omega)$~\cite{Pozzo11,recoules06} or modified Drude
forms~\cite{JunYan-Be14}). More discussion of these issues is given in the
Appendix. The $N$-ion DFT+MD model does not
allow an easy estimate of single-ion properties, e.g., the mean number of free
electrons per ion ($\bar{Z}$) or ion-ion pair potentials.

The NPA methods, e.g., that of Perrot and Dharma-wardana, reduce
the many-electron, many-ion problem to an effective one-electron, one-ion
problem using DFT~\cite{DWP1,eos95,CPP}. A Kohn-Sham (KS) calculation for a
nucleus immersed in the plasma medium provides the  bound and
free KS states. While bound states remain localized within the Wigner-Seitz
(WS) sphere of the ion for the regime studied here, the free electron
distribution $n_f(r)$ of each ion resides in a large ``correlation sphere''
(CS) such that all $g_{ij}(r)\to$ 1 as $r \to R_c$. We typically use 
$R_c=10r_{ws}$, i.e., a volume of some 1000 atoms.  Several average-atom
models~\cite{Murillo13,Blenski13} have similarities and significant differences
among them and with the NPA method. These are reviewed in the Appendix and in 
Ref.~\cite{cdw-carbon16}.  The NPA
method applies for low $T$ systems even with transient covalent  bonding. 
Hence, we differ from
Blenski {\it et al.}~\cite{Blenski13} who hold that ``... all quantum
models seem to give unrealistic
description of atoms in plasma at low $T$ and high plasma densities''.
But in reality, the earliest successful applications of the NPA were for
 solids at $T=0$.
Here we treat very low-$T$ WDMs, e.g., Al, $\rho$=2.7 g/cm$^3$,
 $T/E_F<0.01$, using the NPA,
$E_F$  being the Fermi energy and obtain very good agreement for equations of 
state (EOS) data~\cite{HarbPhon17}
and even for transport properties, e.g., the electrical conductivity. 

The NPA static conductivity is evaluated from the Ziman formula using
the NPA pseudopotential $U_{ei}(k)$ and  the ion structure factor
$S(k)$~\cite{eos95} generated from the NPA pair potential $V_{ii}(r)$. The
latter is used in  the hypernetted-chain (HNC) equation or its modified (MHNC)
form inclusive of bridge functions, assuming spherical symmetry appropriate to
fluids. HNC methods are accurate, fast and much cheaper than MD methods which
fail to provide small $k$-information, i.e., less than $\sim 1/L_{bx}$ where $L_{bx}$
 is the linear
dimension of the simulation box. The Ziman formula can be derived from the
 Kubo formula using the force-force correlation function and assuming
  a momentum relaxation time $\tau_{\rm mr}$. Zubarev's method can also
 be used~\cite{Reinholz2000} to derive the Ziman formula. Details regarding
 the conductance formulae and their limitations are given in the Appendix.

\section{The conductivities of WDM aluminum}

Surprisingly low static conductivities for UFM aluminum at 2.7
g/cm$^3$, extracted from x-ray scattering data from the Linac Coherent Light Source (LCLS)
have been reported in Sperling {\it et al.}, Ref.~\cite{Sper15}. 
Calculations of $\sigma_{\rm ic}$ using an orbital-free
(OF) form of DFT and MD revealed sharp 
disagreement with the LCLS data~\cite{SjosCond}. 
Sperling {\it et al.}~\cite{Sper15} found the conductivity data of Gathers~\cite{IJTher83} 
to differs strikingly from the LCLS data and the OF results. 
In Fig.\ 4 of Ref.~\cite{Sper15}, they attempt to  present a
theoretical $\sigma_{\rm ic}$ at 2.7 g/cm$^3$ that agrees approximately
 with the Gathers'  data and to some extent with the LCLS data.
The Gathers data are reviewed in the Appendix.  

\begin{figure}[t]
\includegraphics[width=\columnwidth]{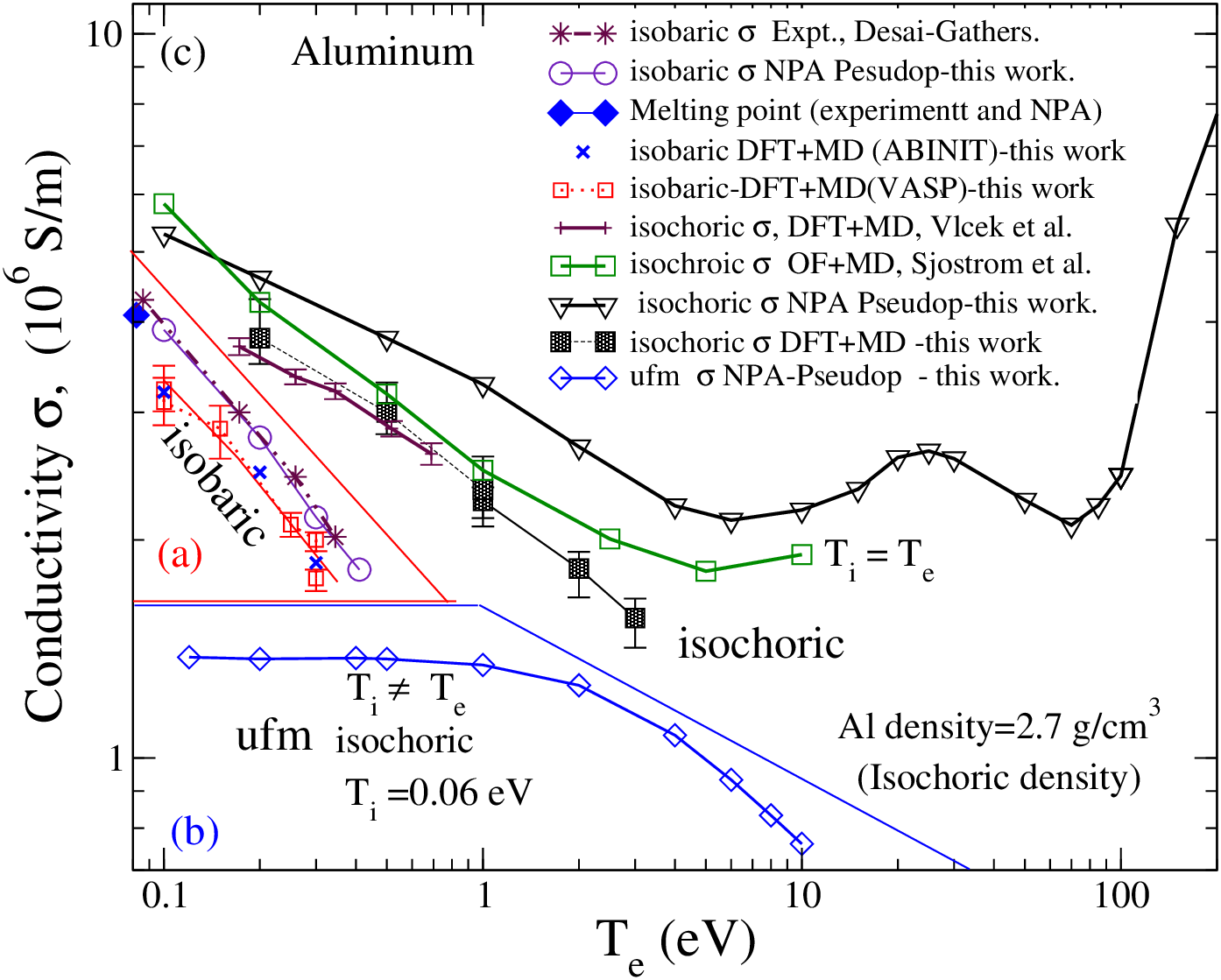}
 \caption{(Color online) 
Static conductivities for Al from experiment and from DFT+MD and NPA calculations. 
The isobaric conductivity
$\sigma_{\rm ib}$ is at densities $2.37\le\rho\le 1.65$ g/cm$^3$ (cf.
triangular region (a)). The  isochoric ($\sigma_{\rm ic}$, region (c)) and UFM
($\sigma_{\rm uf}$, region (b)) conductivities are  for a density of 2.7
g/cm$^3$. Enlarged views of regions (a) and (c) are given in the Appendix.
The blue-filled diamond gives the  conductivity of normal aluminum
at its melting point (0.082 eV, 2.375 g/cm$^3$), viz.\
$\sigma=4.16\times10^6$ S/m from experiment (quoted in 
Ref.~\cite{LeavensPhysChemLiq81}) and $\sigma=4.09\times10^6$ S/m from the NPA.
}
\label{cond-fig}
\end{figure}

However, in our view, the LCLS, OF, and
 Gathers $\sigma$ {\it should} indeed differ,  in
the physics involved as well as in the actual values,
because:\\
 (i) the Gathers data are for the {\it isobaric} conductivity
$\sigma_{\rm ib}$ of liquid aluminum from $\rho=1.7$ to 2.4 g/cm$^3$ (cf.\ 
region (a) in Fig.~\ref{cond-fig}).\\
 (ii) The orbital-free simulation~\cite{SjosCond} and the DFT+MD
 simulations~\cite{Vlcek12} are
for the {\it isochoric} equilibrium ($T_e=T_i$)  $\sigma_{\rm ic}$ of Al at
$\rho$=2.7 g/cm$^3$ (region (c) in Fig.~\ref{cond-fig}).\\
 (iii) The  LCLS
data applies to UFM-aluminum $\sigma_{\rm uf}$, $T_i\ne T_e$, with the ions
`frozen' at $T_i\simeq T_0$, as  proposed in Ref.~\cite{cdw-plasmon16}. The
UFM-conductivity is shown as region (b) in Fig.~\ref{cond-fig}. The ultrafast
conductivity $\sigma_{\rm uf}$ is essentially isochoric, with $T_i\simeq T_0$
at the density $\rho_0$. The timescales in UFM experiments are too short for
$(\rho, T_i)$ to differ significantly from $(\rho_0, T_0)$. The evaluation of the
 ultrafast conductivity $\sigma_{\rm uf}$ was discussed in detail
 in Ref.~\cite{cdw-plasmon16},
and here we extend our study of ufm-conductivities.

 
\begin{figure}[t]
\includegraphics[width=\columnwidth]{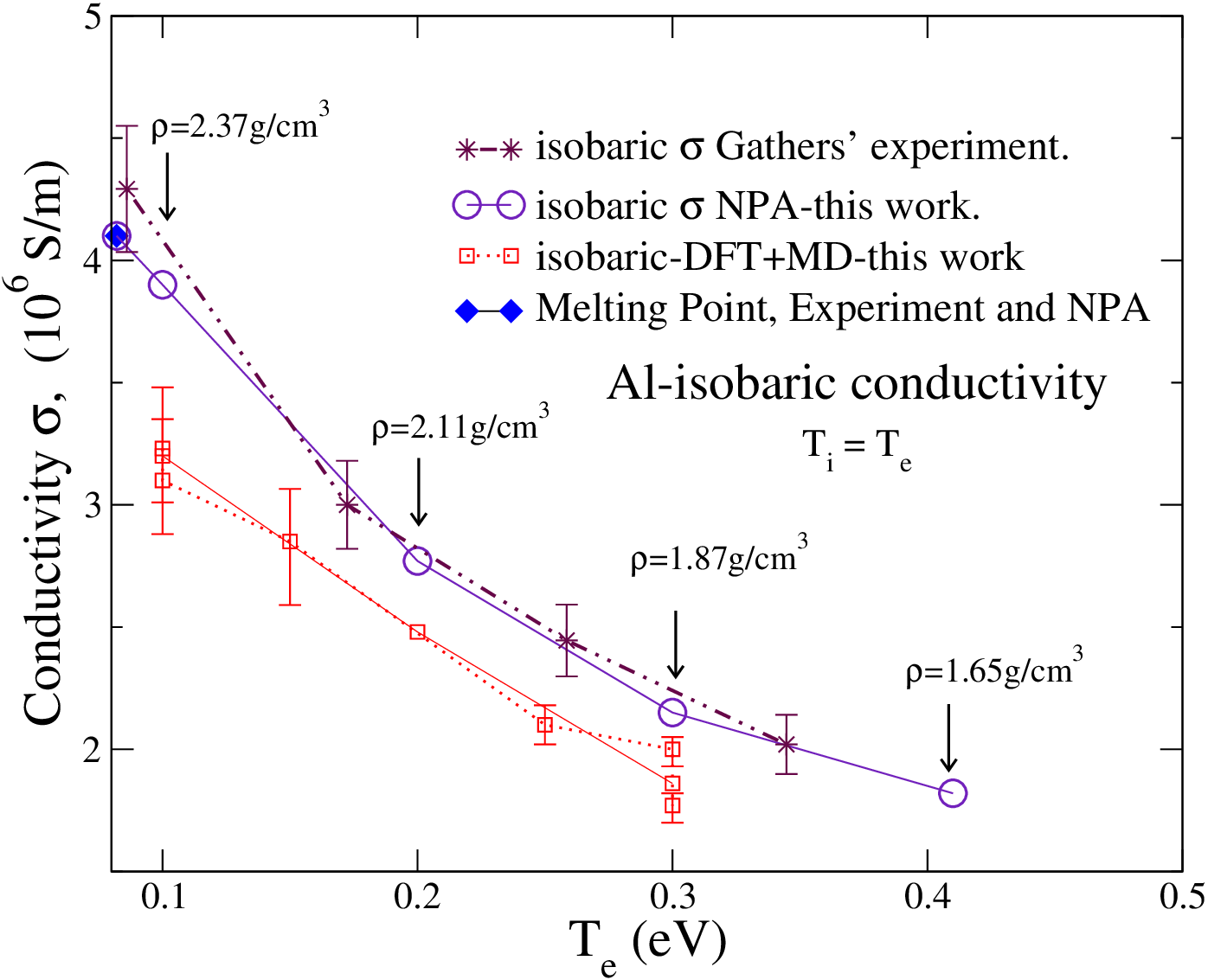}
\caption{(Color online) Isobaric conductivity of aluminum from near its melting
point to about 0.4 eV, expanded from Fig.\ 1  
comparing the
NPA, experiment (Gathers) and DFT+MD results. The experimental conductivity of Al
at its melting point (filled blue diamond)~\cite{LeavensPhysChemLiq81},
 with density 2.375 g/cm$^3$ is displayed and aligns with the
NPA calculations for the Gathers data showing the very good agreement between the
NPA and two independent experiments.} 
\label{al-sig-isob.fig}
\end{figure}

\subsection{Isobaric conductivity}
In Fig.~\ref{cond-fig},  we globally compare our  NPA-Ziman {\it isobaric}
conductivities for aluminum with the isochoric and ufm conductivities,
shown in regions (b) and (c). The three conductivities 
evolve in characteristic ways as a function of temperature.

The experimental data of Gathers for $\sigma_{\rm ic}$ are compared with our results in
more detail in Fig.~\ref{al-sig-isob.fig}, and we find {\it
excellent agreement} with our NPA calculation.  DFT+MD calculations
using a 108-atom  
simulation cell are shown in both figures for $\sigma_{\rm ib}$ and $\sigma_{\rm ic}$,  where
the PBE functional available in VASP and ABINIT was used; they fall below the experimental
$\sigma_{\rm ib}$ or the NPA $\sigma_{\rm ib}$, a common trend  for the DFT+MD+KG  $\sigma_{\rm ic}$
as well, as discussed further in the Appendix. It should be noted that Gathers gives
two isobaric resistivities in columns four and five of Table-II (Ref.~\cite{IJTher83}),
causing some confusion;  Gathers' results are further discussed
in the Appendix.

The {\it isochoric} conductivity of Al at 0.082 eV (nominal melting point)
 is $\approx 5\times 10^6$ S/m;
the experimental {\it isobaric} conductivity~\cite{LeavensPhysChemLiq81} at the
melting point is
$\sigma_{\rm ib}=4.1\times 10^6$ S/m, with density
 2.375 g/cm$^3$  instead of the room-temperature density of 2.7 g/cm$^3$
 due to thermal expansion.
The value of 4.08 $\times 10^6$ S/m obtained from NPA for aluminum at 2.375 g/cm$^3$  is
 in excellent agreement with experiment. It is
shown as a filled blue diamond symbol in figure~\ref{cond-fig}. This value drops to
 3.8 $\times 10^6$ S/m if a bridge contribution (MHNC) is not used
in calculating the ion-ion structure factor.

\subsection{Isochoric conductivity}
The isochoric system, region (c) in Fig.~\ref{cond-fig},
 is at $\rho_0=$ 2.7 g/cm$^3$,
 $r_{ws}\simeq 2.98$ a.u. ($\hbar=|e|=m_e=1$),  for all $T=T_i=T_e$.
 The NPA value of $\sigma_{\rm ic}$
 at $T=0.082$ eV (nominal melting point) is $\approx 5 \times 10^6$ S/m;
 this is higher than the experimental value usually quoted \cite{LeavensPhysChemLiq81} of 
$\sigma_{\rm ib}=4.1\times 10^6$ S/m as the density of normal aluminum
becomes 2.375 g/cm$^3$  instead of 2.7 g/cm$^3$ due to thermal expansion.
In region (c) we see the OF conductivity of Ref.~\cite{SjosCond} going to a
  minimum at $T\sim$5 eV  and  
 subsequently rise as $T$ increases; 
 DFT+MD+KG becomes increasingly prohibitive at these higher
 temperatures. The
NPA calculations show a first minimum at $\sim$ 6 eV, followed by a maximum at 25 eV, and
another minimum at $\sim$ 70 eV. These features in the NPA results
are  due to the concurrent increase in
$\bar{Z}$ as well as the competition between different ionization states. 
This effect --- the conductance minimum or resistivity saturation --- occurs when 
electrons become non-degenerate (i.e., $\mu_e\le 0$), i.e. when all
electrons (not just those  near $E_F\sim12$ eV) begin
to conduct. 

While we favour this explanation of the minimum in the
conductivity and first presented it in our discussion~\cite{Thermophys99}
of the Mlischberg experiment, some authors (e.g., R. M. More in Ref.~\cite{Milchberg88},
and also Faussurier {et al.}~\cite{Faussurier15}) have proposed an explanation
in terms of resistivity saturation, as in Mott's theory of minimum
conductance in semiconductors. The electron ``mean-free path''
 $\lambda=\bar{v}\tau_{\rm mr}$,
where $\bar{v}$ is a mean electron velocity,  is claimed  to
reduce to the mean interatomic distance at resistivity saturation. However,
 $\tau_{\rm mr}$ evaluated using the
Ziman formula  is a momentum-relaxation
time associated with scattering within the thermal window  of the Fermi distribution
at the Fermi energy (more accurately, at an energy corresponding to the chemical potential).
Since $2k_F$ is of the order of an inverse $r_{ws}$, it is not surprising that
one can connect a length scale related to $r_{ws}$ to $\lambda$. But it
does not describe the right physics of the conductivity minimum. Even the simplest form
of the Ziman formula already shows the conductivity minimum, and it is a single-center scattering
formula using a Born approximation within a continuum model; it contains no information on
 the interatomic distance since one can even set $S(k)=1$ and obtain the resistivity saturation. 
In contrast, the resistivity saturation seen in Fig.~\ref{cond-fig} for the NPA calculation
manifests itself from
approximately half the  Fermi energy ($\simeq$ 6 eV) corresponding to $\bar{Z}=3$, to about 70 eV
 corresponding to a much higher ionization of $\bar{Z}\approx 7$. The increased ionization
prevents the chemical potential from becoming rapidly negative and delays the onset of the 
steep rise in conductivity. These features cannot be explained via
 a limiting mean free-path model. In fact, in an isochoric system the interionic
distance does not change and one cannot have the complex structure shown in the
NPA $\sigma_{\rm ic}$ in such a model.  For $T_e=6$ eV to about $T=25$ eV,  $\bar{Z}=3$ for
Al and steadily converts to $\bar{Z}=4$, and then a decline and a rise are
accompanied by the conversion of $\bar{Z}=4$ to $\bar{Z}\simeq$7 by  $T\sim70$ eV.    

Fig.\ 1 of Faussurier {\it et al.}~\cite{Faussurier15} displays the isochoric
 resistivity for aluminum together with results from Perrot and
 Dharma-wardana~\cite{Thermophys99}. However, the latter gives the scattering
 as well as the pseudopotential-based resistivity for aluminum where the
 mean {\it electron} density $\bar{n}$
is held constant, not the usual isochoric resistivity where the ion density
$\bar{\rho}$ has to be held constant. Electron-isochoric  and ion-isochoric
 conditions are equivalent
 initially and as long as $\bar{Z}=3$ for aluminum; but the comparison becomes
 misleading beyond $T\approx15$ eV. 
Fig.\ 1 of Faussurier {\it et al.}~\cite{Faussurier15} also displays the
 aluminum isochoric
 resistivity  from Yuan {\it et al.}~\cite{Yuan96}. However, as explained in
 sec.~\ref{diffs-npa.sec}
of the Appendix, both Faussurier and Yuan use an ion-sphere model which leads
 to ambiguities
in the definition of $\bar{Z}$ and $\mu_0$, leading to non-DFT features
 which are absent
 in the NPA model. Hence their resistivity estimates  are not directly
 comparable to ours. 
Sufficiently accurate experiments are not yet available at such high
 temperatures to
 distinguish between different theories and validate one or the other.
 Such models should also be
tested using cases where accurate experimental data are available
 (e.g., in the liquid-metal regime).

A further aspect of conductivity calculations is the  need to account for
multiply-ionized species. For $T>E_F/2$,  $\bar{Z}$ begins to increase beyond  3
and departs substantially from an integer (e.g., $\bar{Z}=$ 3.5 at 20 eV ).
 It is thus clear that a multiple ionization model
with several integral values of $\bar{Z}$, (e.g, a mixture $\bar{Z}=3$ and $\bar{Z}=4$)
 should be used, as implemented in 1995 by  Perrot and Dharma-wardana~\cite{eos95},
 for lower-density aluminum.
 The isochoric data $\sigma_{\rm ic}$  reported
in Fig.~\ref{cond-fig} uses the approximation of a  single ionic species with a
 mean $\bar{Z}$.

\subsection{Ultra-fast conductivity}

 The nature of ultrafast matter and its properties
are determined by the initial state of the system. That is, if the initial system were a room temperature
solid, and if the experiment were performed with minimal delay after the pump pulse 
of the laser, then the ion subsystem would remain more or less intact. However, the initial state
can also be the liquid state and this will lead to  different results.
Both these cases are studied to compare and contrast the resulting $\sigma_{\rm uf}$ for Al.

\begin{figure}[t]
\includegraphics[width=\columnwidth]{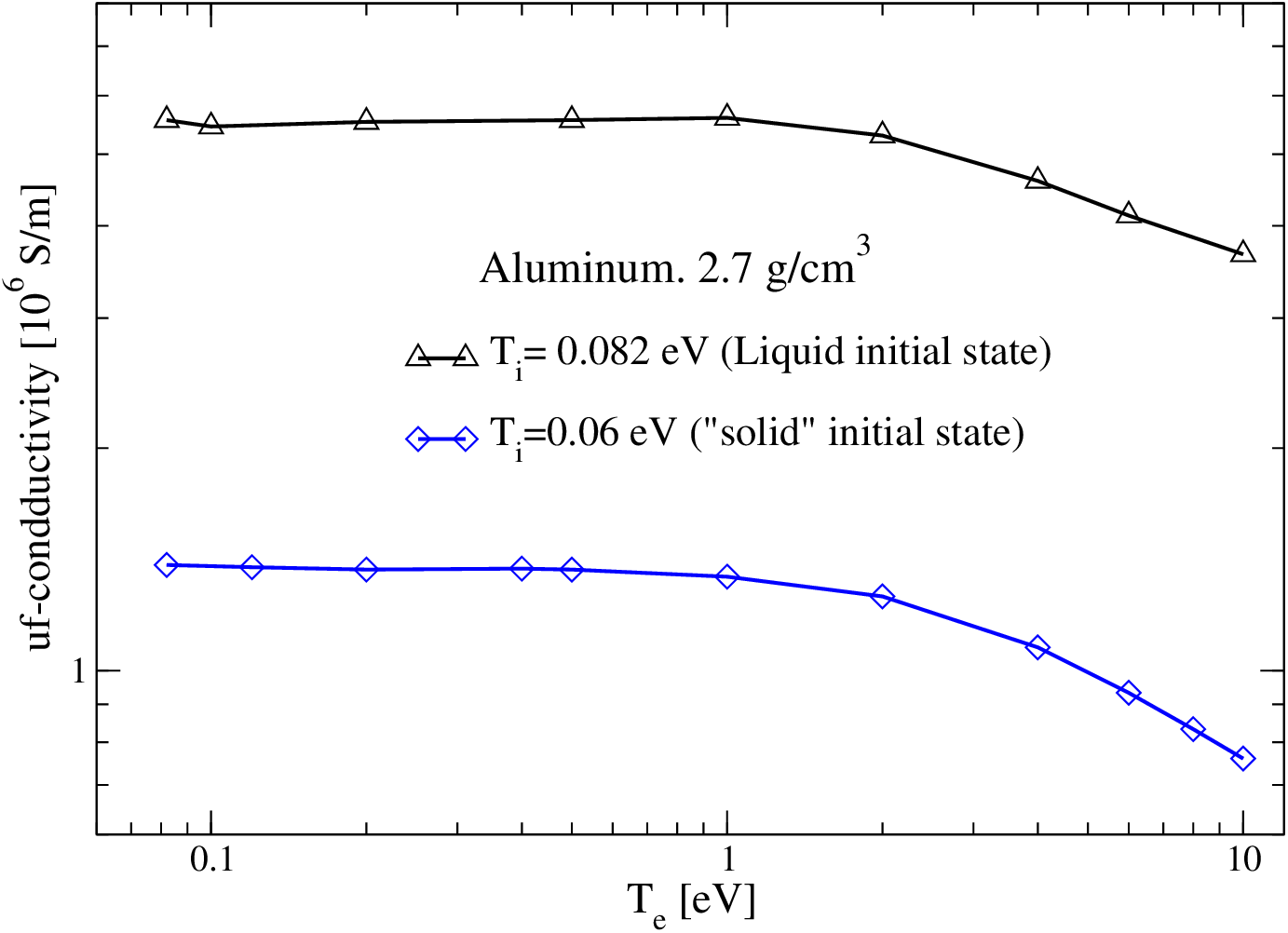}
\caption{(Color online) Ultra-fast conductivity of Al at
density 2.7 g/cm$^3$ for (i) solid initial state at 0.06 eV, and 
 (ii) liquid initial state just above melting point 0.082 eV.
Curve (i) was also displayed in Fig.~\ref{cond-fig} for comparison with
$\sigma_{ib}$ and $\sigma_{ic}$.}
\label{uf2sigma.fig}
\end{figure}

(i) For the case where the initial state is solid (FCC lattice),
 we assume for simplicity
that the ion subsystem structure factor $S(k)$ can be adequately approximated
 by its spherical average
since aluminum is a cubic  crystal. The major Bragg contributions are
included in such an approximation. In fact, the spherically-averaged
 $S(k)$ is taken to
be the ion-ion $S(k)$ of the supercooled liquid at 0.06 eV 
as that is the lowest temperature (closest to room temperature) where
the Al-Al $S(k)$ could be calculated. The results are in fact
insensitive to whether we use the $S(k)$ at 0.06 eV, 0.082 eV (melting point)
 or 0.1 eV.
Furthermore, here we are using the simplest local
($s$-wave) pseudopotential derived from the NPA approach using a radial KS equation.
Hence the use of a spherically average $S(k)$ is consistent, and probably within the
large error bars of current LCLS experiments (see Fig.4, Ref.~\cite{Sper15}).
The NPA $\sigma_{\rm uf}$ results (Fig.~\ref{uf2sigma.fig}) for the case  where the initial
 state is below the melting point (mimicking solid Al) have been compared in detail with
the experimental data in Ref.~\cite{cdw-plasmon16}.
 Currently, no DFT+MD+KG results for
 $\sigma_{\rm uf}$ are available for comparison.  One notes that 
 the $\sigma_{\rm uf}$ at $T_e=T_i=0.6$ eV does not go to the conductivity of
solid (crystalline) aluminum, but goes to a  lower value, possibly consistent
 with that of a supercooled liquid. The lower conductivity,
 compared to the FCC crystal is
qualitatively consistent with the drop in the conductivity from the solid-state, 
room temperature, density = 2.7 g/cm$^3$ 
value of  $\sigma\simeq 41\times 10^6$ S/m
 to the liquid-state
value  at the melting point, 4 $\times 10^6$ S/m.
 The drop predicted by
the NPA $\sigma_{\rm uf}$ is larger. Hence this calculation appears to need
further improvement for $T< 1.0$ eV, e.g.
using the  structure factor of the FCC solid and including
appropriate band-structure effects.

(ii) The second
model we study has molten Al at its nominal melting point (0.082 eV) 
but at its isochoric density of 2.7 g/cm$^3$ as the initial state. This mimics the case where the pump
pulse had warmed the ion subsystem to some extent. Then we use the $S(k)$ 
and  pseudopotentials evaluated at 0.082 eV (nominal melting point), and regard
that they remain unchanged while the electron
screening and all properties dependent on the electron subsystem are evaluated at the 
electron temperature $T_e$. The resulting $\sigma_{\rm uf}$ is shown in Fig.~\ref{uf2sigma.fig},
together with the case where the initial state was assumed to be a temperature (e.g.,the
 room temperature 0.026eV, or 0.06 eV)
which is below the melting point. The two curves clearly suggest that the LCLS-experiments
(see Ref.~\cite{cdw-plasmon16} for details)  for Al are  more consistent with
the initial state (i.e., the state of matter at the peak of the  laser pulse) being 
solid, and with no significant pre-melt.

\subsection{XC-functionals and the Al conductivity}
Using a DFT+MD+KG approach, Witte {\it et al.}~\cite{WittePRL17} examined the
$\sigma$ for Al at $\rho =$ 2.7 g/cm$^3$ and  $T =$ 0.3 eV computed with the
exchange-correlation (XC) functionals  of (i) Perdew, Burke, and Ernzerhof
(PBE) \cite{PBE96} and (ii) Heyd, Scuseria, and Ernzerhof (HSE) \cite{HSE03}.
Their results agree with those of Vl\v{c}ek {\it et al.} \cite{Vlcek12} for the PBE functional; 
our DFT+MD calculations also agree well with those of Vl\v{c}ek {\it
et al.} as seen from the region (c) in Fig.~\ref{cond-fig}. However, Witte {\it et
al.} propose, from their Fig.\ 1, that their HSE calculation agrees~\cite{RedmerPrivate} best with the
experimental data of Gathers~\cite{IJTher83}. This is based on a  calculation of the 
conductivity at 0.3 eV only ($\approx 3500$ K), which is compared with the corresponding entry
in Table II, column 4, of Ref.\ \cite{IJTher83}, viz.\ resistivity=0.451$\mu\Omega$m, i.e., 
conductivity = 2.22 $\times 10^6$ S/m. However, this datum
 is given by Gathers for a volume
dilation of 1.44 (column 3), i.e., $\rho = 1.875$ g/cm$^3$, and \textit{not}
$2.7$ g/cm$^3$. 
Witte {\it et al.} incorrectly interprets 
 column 4 of Gathers' Table II as providing
 {\it isochoric} conductivities of Al at 2.7 g/cm$^3$.
Gathers' tabulation and the several resistivities given
are indeed a bit confusing; we 
reconstruct them in Table~1 of the Appendix for convenience.

Columns 4 and 5 in Ref.\ \cite{IJTher83}  give two possible
results for the isobaric  conductivity of aluminum, with column 5 giving
 the experimental resistivity as a function of
 the  nominal input enthalpy, i.e., ``raw data''.  
Column 4 gives the resistivity where in effect the input enthalpy
  has been corrected  for volume expansion; this 
is {\em not} the isochoric resistivity of aluminum, as proposed by
 Sperling {\it et al.}~\cite{Sper15} and  by
Witte {\it et al.} \cite{WittePRL17}. 

All the resistivities in Gathers’ Table II, column 4 
 can be recovered accurately by our parameter-free NPA
calculation using the isobaric densities. Also, the fit formula given in the last
 row of table 23 of
 Gathers' 1986 review ~\cite{Gathers86}
confirms that Table II, column 4 in Ref.\ \cite{IJTher83} is
indeed the final {\it isobaric} data at 0.3 GPa. 
Our NPA calculation at the melting  
recovers the known isobaric conductivity~\cite{LeavensPhysChemLiq81} at 0.082 eV. 
which is also consistent with the Gathers data.

The HSE functional  includes  
a contribution (e.g., 25\%) of the Hartree-Fock exchange functional in it.
 If there is no band gap at the Fermi energy, the Hartree-Fock self-energy is such that
several Fermi-liquid parameters become singular. Hence the use of this
 functional in WDM studies may lead to uncontrolled or unknown
errors. Furthermore, previous studies, e.g., Pozzo {\it et al.}~\cite{Pozzo11},
Kietzmann {\it et al.}~\cite{Kietz08}, show that the PBE functional
successfully predicts conductivities.
Those conductivities, if recalculated with the HSE functional are most likely
to be in serious disagreement with the experimental data.

DFT is a theory which states that the  free energy is a functional of 
the one-body electron density, and that the free energy is minimized by
just the physical density. It does {\it not} claim to
give, say,  the one-electron excitation spectrum or the density of states (DOS).
 The spectrum
and the DOS are those of a fictitious non-interacting electron system at the
 {\it interacting density},
and moving in the KS potential of the system. The KS potential is not
a mean-field approximation to the many-body potential, but a potential that
gives the exact physical one-electron density if the XC-functional is exact.
Hence any claimed ``agreement'' between the DFT spectra and physical
spectra is not relevant to the quality of the XC-functional, except
in phenomenological theories which aim to go beyond DFT and recover
spectra, DOS, bandgaps etc.,  by including
parameters in `metafunctionals' which are fitted to a wide array of properties.
There is however no theoretical basis for the existence of XC-functionals
which also simultaneously render accurate excitation spectra, DOS and bandgaps
in a direct calculation.

\subsection{The variation of the conductivity as a function of temperature}
\label{T-variation.sec}
The evolution with temperature of the conductivity can be understood within the
physical picture of electrons near the Fermi energy (chemical potential) 
undergoing scattering from the ions in a correlated way via the structure factor.
This in turn invokes  the relation of the structure
factor to the Fermi momentum $k_F$, and  the breakdown of the Fermi
surface as $T/E_F$  is increased, while 
the breakdown is countered by
ionization which increases the Fermi energy. At sufficiently high temperatures
the chemical potential $\mu$ tends to zero and  to negative values. 
The conductivity then
becomes classical, and finally Spitzer-like. The conductivity minimum
 (resistivity plateau) in WDM systems occurs
near the $\mu\approx 0$ region and is not related to the Mott minimum
conductivity.

\begin{figure}[t]
\includegraphics[width=\columnwidth]{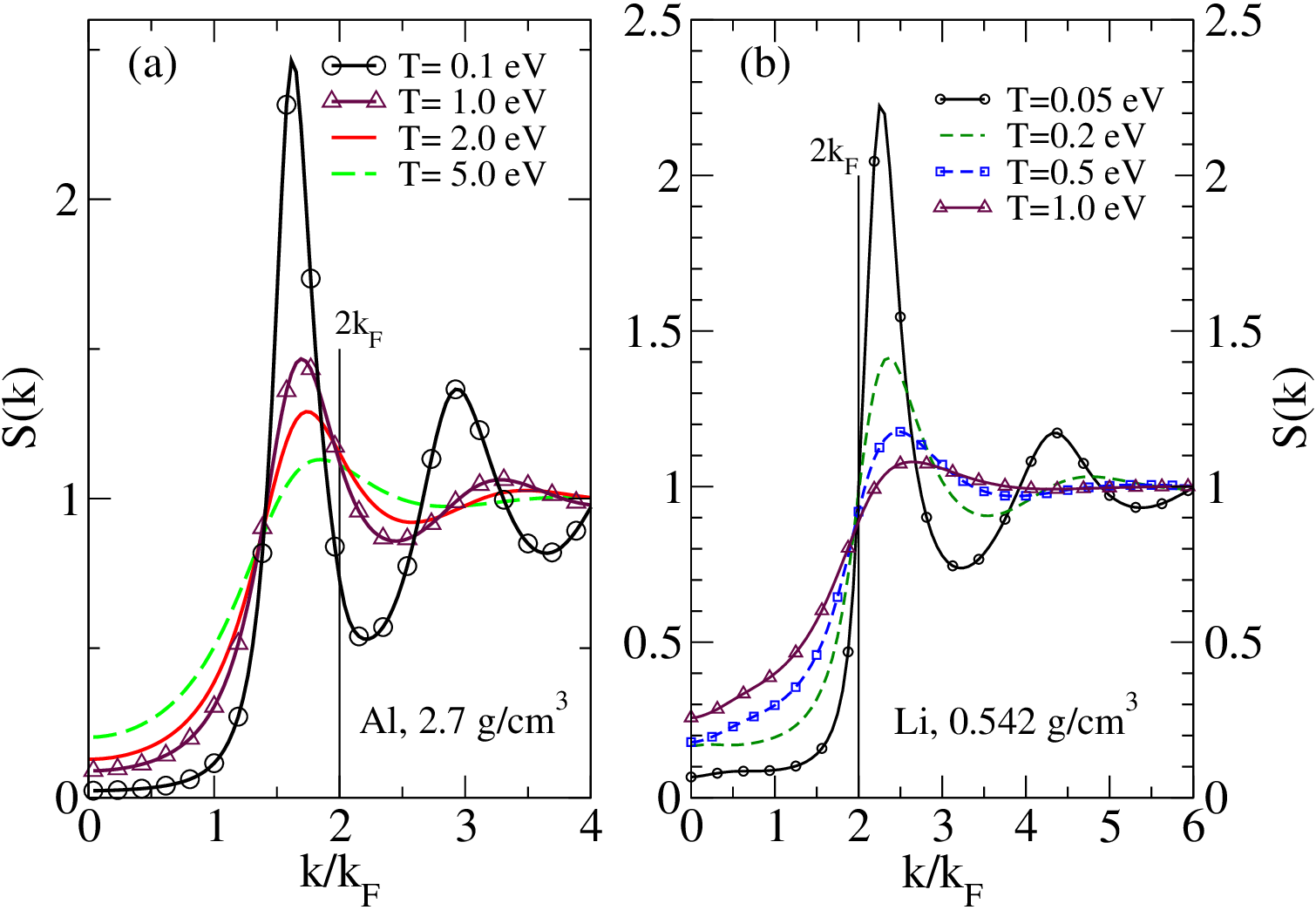}
 \caption{(Color online) (a) Static structure factor $S(k)$ of isochoric
aluminum WDM at different temperatures;  $S(k)$ at 2$k_F$ changes by 65\% from $T=0.1$ to $T=5$ eV.
 In ultrafast aluminum $S(k)$ remains `fixed' at
the initial temperature, even when $T_e$ changes. (b) Evolution of $S(k)$ for
isochoric Li at 0.542 g/cm$^3$ as a function of temperature. 
As $T$ increases, the peak
broadens and shifts away from 2$k_F$.}
\label{sk.fig}
\end{figure}

The differences between $\sigma_{\rm ic}$ and $\sigma_{\rm uf}$, both
isochoric, arise because the structure factors $S(k,T_i)$ of the two  systems are different,
while $U_{ei}(k)$ and the Fermi-surface smearing for them are essentially the
same at $T_e$, with $\bar{Z}\simeq3$ for Al. The ion structure factor at different temperatures,
calculated using the NPA pseudopotential $U_{ei}(k,T_e)$ and  used for
evaluating $\sigma_{\rm ic}$,  are shown in Fig.~\ref{sk.fig} (a). The $U_{ei}$
and $S(k)$, and hence $\sigma$, are  first-principles quantities determined
entirely from  the NPA-KS calculation. If the initial temperature $T_0$
at the time of creation of the Al-UFM were 0.082 eV (i.e., $\sim$melting
point), then the corresponding $S(k,T_0)$ is used in evaluating $\sigma_{\rm
uf}$ at all $T_e$, together with the $U_{ei}(k,T_e)$. More details of
$\sigma_{\rm uf}$ and comparison with LCLS data may be found in
Ref.~\cite{cdw-plasmon16}. The isobaric system  differs from the
isochoric and ultrafast systems due to  volume expansion. Hence the $S(k)$ and
the $U_{ei}$ are calculated at each `expanded' density.

Degenerate  electrons ($T_e/E_F<1$) scatter from
one edge (e.g., $-k_F$)of the Fermi surface to the opposite edge
 ($k_F$), with a momentum change
$k \simeq 2k_F$ and their scattering contribution essentially
determines $\sigma$. Thus the position of 2$k_F$ with respect
to the main peak of $S(k)$ and its changes with  $T_e$ explain the $T_e$
dependence of $\sigma(T_e)$. For aluminum at $\rho=2.7$ g/cm$^3$, 2$k_F$ lies
on the high-$k$ side of the main peak, and as $T_i=T_e$ increases, the
peak broadens into the 2$k_F$ region (see Fig.~\ref{sk.fig}(a)), resulting in
increased scattering. In the isochoric UFM case both  $T_i$ and $S(k)$ do not change, but
as $T_e$ increases the window of scattering $f(k)(1-f(k))$  increases (here $f(k)$
is the finite-$T$ Fermi occupation number), and $\sigma_{\rm uf}$ decreases.

Given that the NPA is a first-principles (i.e., `parameter-free')  DFT scheme,
the excellent agreement between the NPA $\sigma_{\rm ib}$ and the Gathers
aluminum  data for $\sigma_{\rm ib}$ (see Fig.~\ref{al-sig-isob.fig})
 confirms the accuracy of NPA
 pseudopotentials $U_{ei}$ and
structure factors, and enhances our confidence in
 the NPA predictions for $\sigma_{\rm ic}$. In addition,
 experiments at other density ranges were found to be in
 good agreement with NPA 
calculations~\cite{Benage99} and with the
DFT+MD calculations of Dejarlais {\it et al.}~\cite{Dejarlais02}. Furthermore,
 the NPA approach becomes
more reliable at higher temperatures ($T/E_F>1$) while the DFT+MD methods rapidly
 become impractical due to the large number of electronic states that are needed in the
 calculation due to the spread in the Fermi distribution. 
At lower $T$  ion-ion correlations and interactions become important and DFT+MD treats them
 well. However, at low $T$, the higher conductivities imply longer mean free paths
 and the need for  simulation
cells with larger $L_{bx}$~\cite{DarioPrivate17}. Good  DFT+MD+KG results, when
 available, provide  benchmarks for calibrating other methods.    

\begin{figure}[t]
\includegraphics[width=\columnwidth]{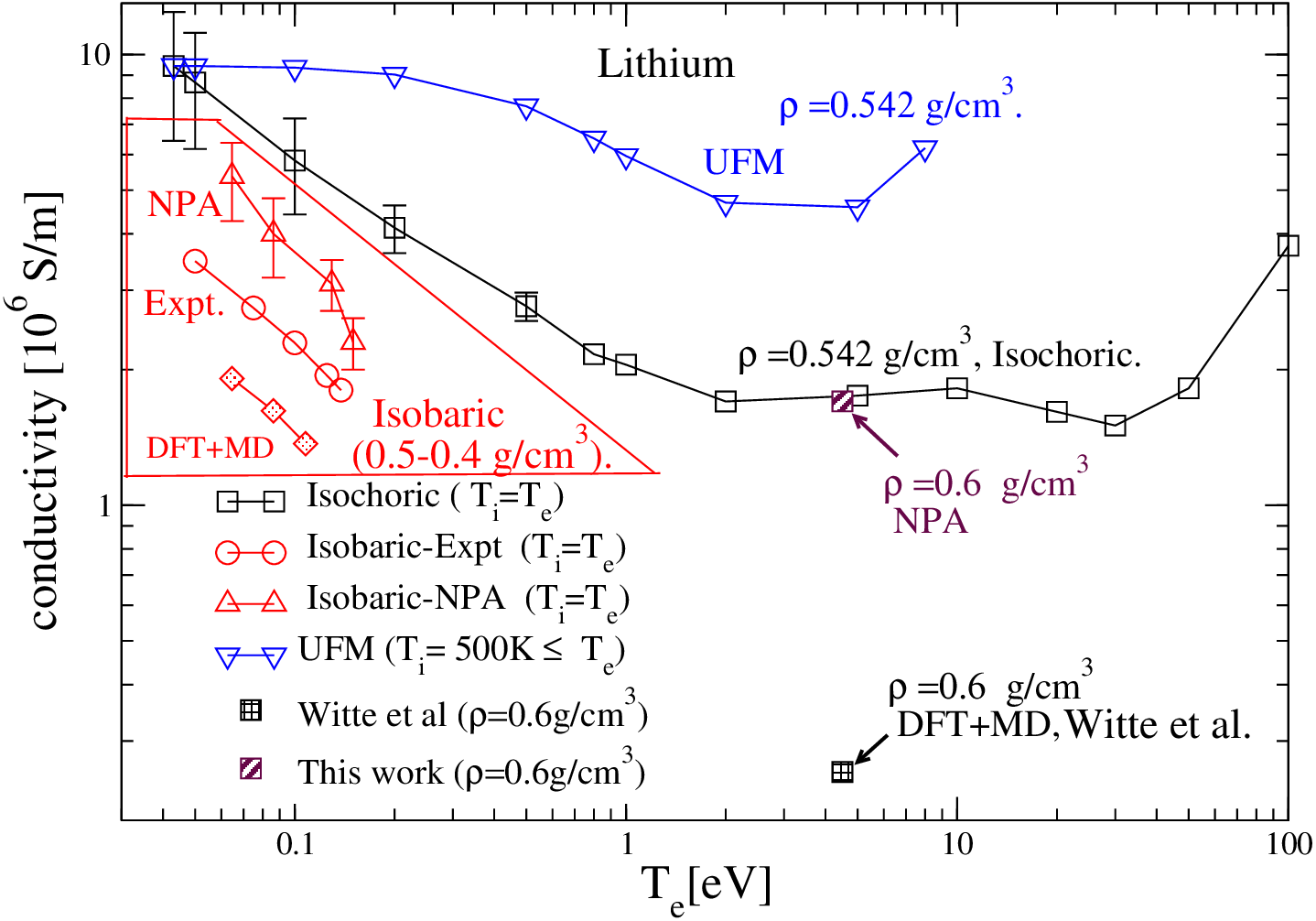}
\caption{(Color online) Isobaric ($\sigma_{\rm ib}$), isochoric($\sigma_{\rm ic}$),
and ultrafast ($\sigma_{\rm uf}$) conductivities of Li at density 0.542 g/cm$^3$.
Isobaric experimental conductivities $\sigma_{\rm ib}$ are for $0.5\le\rho\le0.4$ g/cm$^3$.
The DFT+MD+KG $\sigma_{\rm ic}$ value of Witte {\it et al.} at 0.6 g/cm$^3$
  and the NPA-Ziman value for $\sigma_{\rm ic}$ are also shown.}
\label{Licond.fig}
\end{figure}

\section{The conductivities of WDM lithium}
The three conductivities $\sigma_{\rm ic},\sigma_{\rm ib}$, and $\sigma_{\rm uf}$
for Li are shown in Fig.~\ref{Licond.fig}. The isobaric data are in the
triangular region. The isochoric conductivities $\sigma_{\rm ic}$
 at a density of $\rho$=0.542 g/cm$^3$, i.e., $r_{ws}=3.251$, are given for a range of $T$, while
 one value at $\rho$=0.6 g/cm$^3$ and $T_e=T_i=$4.5 eV,
 is also given. This is for conditions reported by Witte {\it et
al.}~\cite{Witte17}. The experimental  isobaric data from Oak Ridge~\cite{ORNL88}
 for $\sigma_{\rm ib}$  (0.5 g/cm$^3$ at 0.05 eV to 0.4
g/cm$^3$ at 0.1378 eV), as well as the NPA $\sigma_{\rm ib}$, are also shown.
Unlike aluminum, Li is a ``low electron-density'' material with $\bar{Z}=1$. 
Hence its $E_F\sim 5$ eV is small compared to that of aluminum. For Li, $2k_F$
lies on the low-$k$ side of the main peak as can be seen in
Fig.~\ref{sk.fig}(b). The UFM conductivity $\sigma_{\rm uf}$ remains higher
than the $\sigma_{\rm ic}$, and its temperature dependence can be
understood,  as discussed in sec.~\ref{T-variation.sec}, by the position
 of $k_F$ with respect to $S(k)$ as $T_e$ varies.

\begin{figure}[t]
\includegraphics[width=\columnwidth]{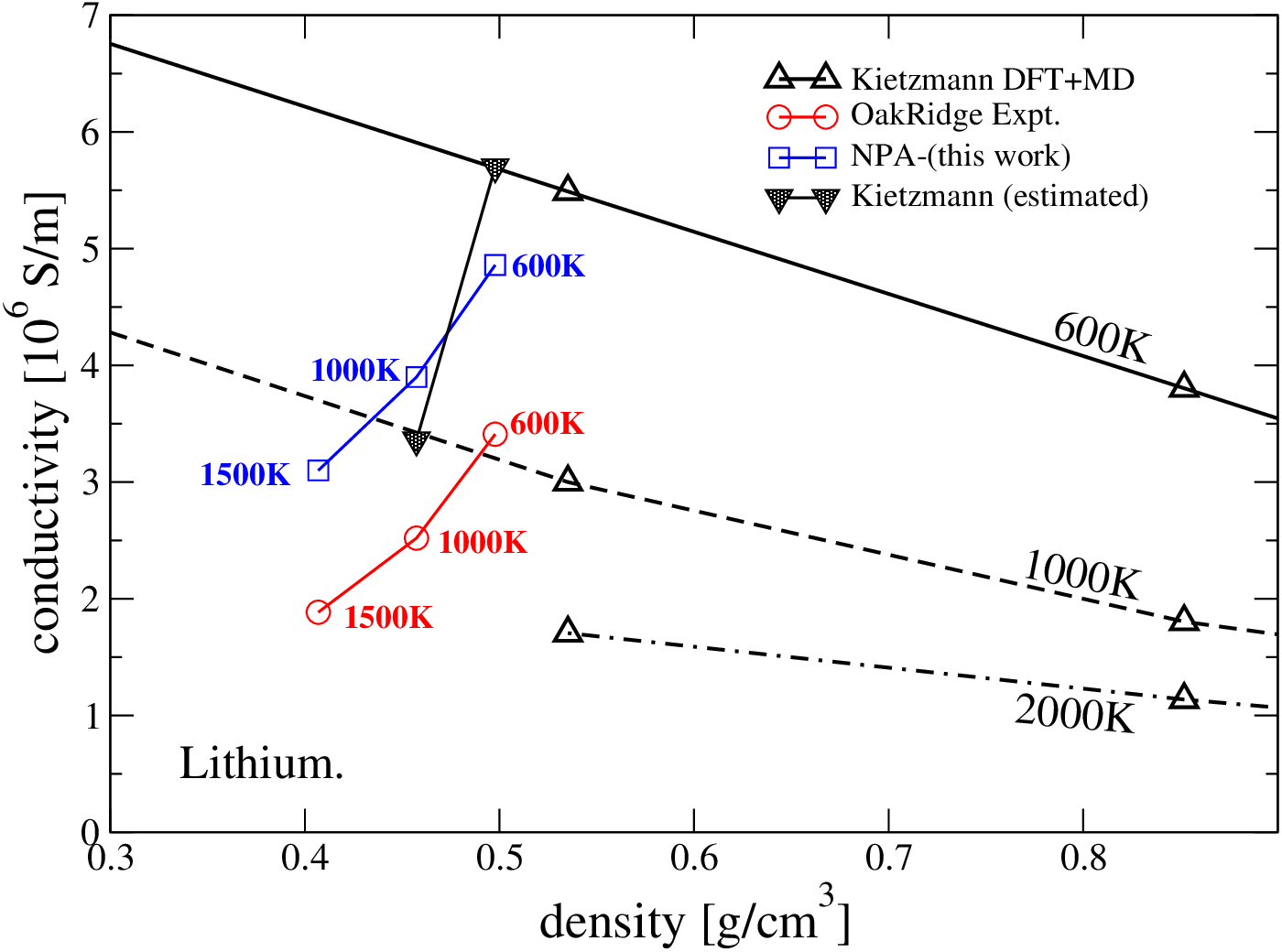}
\caption{(Color online) The Oak Ridge experimental data compared with the NPA and the
DFT+MD+KG conductivity of Kietzmann {\it et al.}~\cite{Kietz08}. Their 600K and 1000K
results have been slightly extrapolated to the low-density region covered by the
experiments. The curve at 2000K given by Kietzmann {\it et al.} is above
 the boiling point of Li, and is not representative of the behaviour of Li at 1500K. }
\label{Li-Kietz.fig}
\end{figure}

The agreement between the NPA-$\sigma_{\rm ib}$ and the Oak Ridge data for
isobaric Li is moderate.  The NPA-Li
pseudopotential is the simplest  local ($s$-wave) form and  corrections
(e.g., for the modified DOS) have not been used. In Fig.~\ref{Li-Kietz.fig}
we have attempted to compare the Oak Ridge experimental data for liquid lithium 
with the DFT+MD+KG calculations of Kietzmann {\it et al.}~\cite{Kietz08}. We use their
calculations as a function of density for 600K and 1500K. The Kietzmann calculation at 2000K
is also shown in Fig.~\ref{Li-Kietz.fig}, but since the boiling point of lithium
under isobaric conditions is $\simeq 1600$K, their calculation at 2000K
cannot be justifiably used to estimate  a value for 1500K from the data of Kietzmann {\it et al.}
which also include the two points at 600K and 1000K.  Nevertheless, their results are consistent
with the observed trend and agree with our NPA results to  the same extent as with the
 Oak Ridge data. 

Disconcertingly, the NPA+Ziman and the DFT+MD  $\sigma_{\rm ic}$ for 
$\rho=$0.6 g/cm$^3$ and $T=4.5$ eV reported by Witte {\it et al.}~\cite{Witte17} using a
64-atom simulation cell disagree by a factor of five.  But the NPA-XRTS calculations
for Li (see the Appendix) agree very well with the DFT-XRTS of Witte {\it et al.}. 
Furthermore, we had already shown that the pair-distribution functions from NPA for
Li for the density range of interest are in good agreement with the simulations of
Kietzmann {\it et al.} (see Ref.~\cite{HarbPhon17}).
However,  at $T$=4.5 eV, $\mu$=0.035 a.u., i.e., the plasma is nearly classical. 
Hence small-$k$ scattering becomes important in determining  $\sigma$.  A simulation cell of
 length $a$=20.26 a.u.holds for 64 atoms. The smallest
momentum accessible is $\pi/a=0.16$/(a.u.), and  fails to capture the smaller-$k$
contributions to  $\sigma$. 
These could  cause the observed differences  between the NPA and DFT+MD+KG
results.

\section{The conductivities of WDM carbon}
Solid carbon is covalently bonded, with strong  $sp3, sp2, sp$ bonding
(with a bond energy of $\sim$
8 eV) being possible.  Hence efforts to create potentials extending to several
neighbors, conjugation and torsional effects etc., have generated  complex
semi-empirical  ``bond-order'' potentials parametrized to fit data bases but
without any $T$ dependence. Transient C-C bonds occur in  liquid-WDM carbon.
Normal-density liquid C near its melting point is a good Fermi liquid with 
four  `free' electrons ($\bar{Z}$ =4) per carbon. An early comparison of
Car-Parrinello calculations for carbon with NPA was reported
by Dharma-wardana and Perrot in 1990~\cite{DWP-Carb90}.
NPA successfully predicts the $S(k)$ and $g(r)$,
inclusive of pre-peaks due to C-C bonding~\cite{cdw-carbon16} as also obtained
from DFT+MD simulations of WDM-carbon\cite{kraus13, Whitley15}. The  NPA 
and  Path Integral Monte Carlo  $g(r)$~\cite{Driver12} also agree
 closely~\cite{cdw-carbon16}.
No experimental $\sigma_{\rm ib}$ are available; hence we  calculate only
$\sigma_{\rm ic}$ and $\sigma_{\rm uf}$ to display the remarkable
difference in the conductivities of  complex WDMs with (transient) covalent
bonding,  compared to simpler WDMs like Al and Li.
\begin{figure}[t]
\includegraphics[width=\columnwidth]{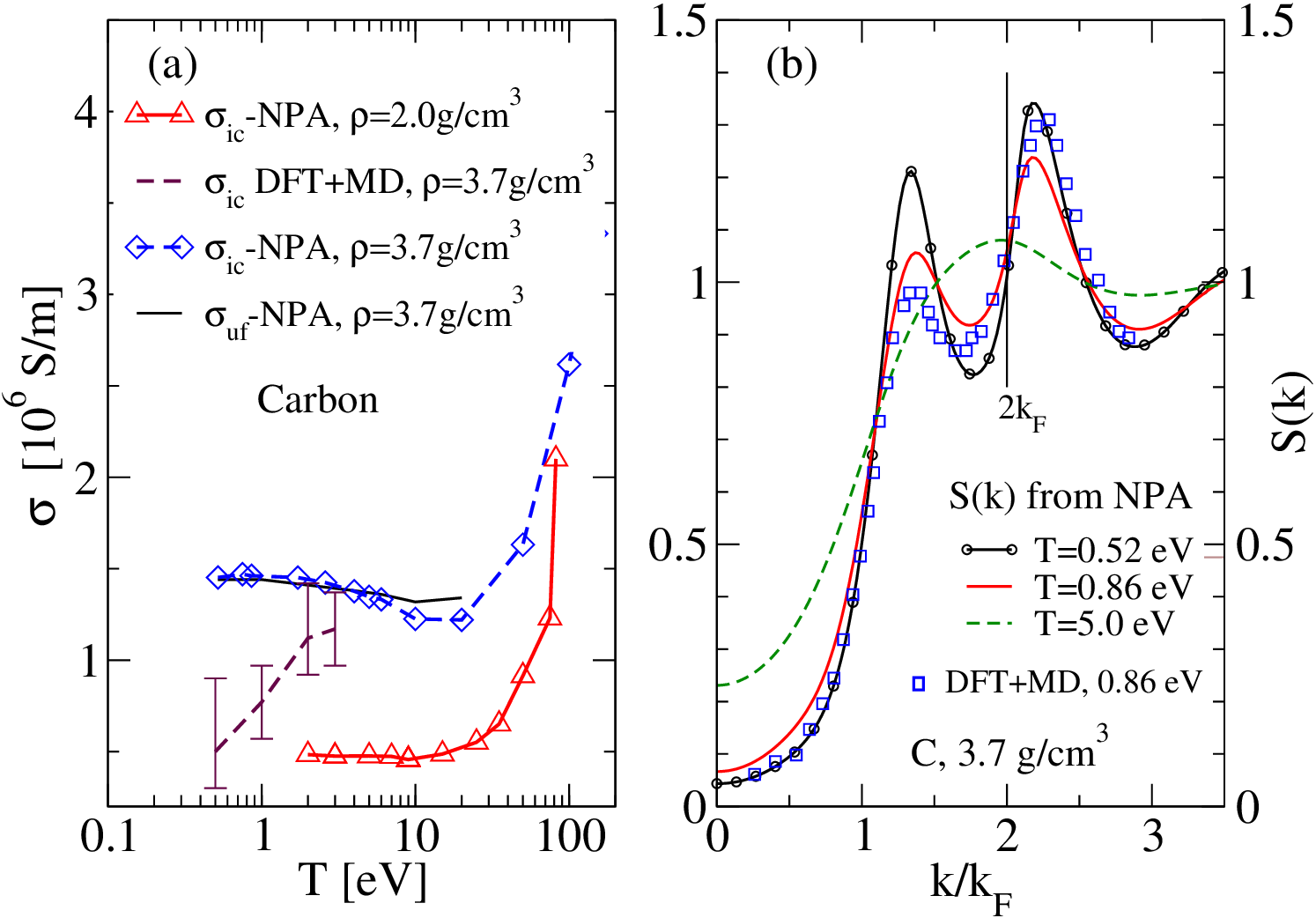}
\caption{(Color online) (a) Isochoric conductivity, $\sigma_{\rm ic}$,  and
ultrafast conductivity $\sigma_{\rm uf}$ for carbon at $\rho =$ 3.7 g/cm$^3$
from NPA and DFT+MD, and isochoric conductivity from NPA for $\rho$=2.0
g/cm$^3$. (b) Ion-ion $S(k)$ for several temperatures; note nearly constant
value of $S(k)$ at 2$k_F$ (indicated by a vertical line).}
\label{carb-cond.fig}
\end{figure}

Figure~\ref{carb-cond.fig}(a) displays $\sigma_{\rm ic}$ and $\sigma_{\rm uf}$ 
for  isochoric carbon at 3.7 g/cm$^3$. Here  $E_F$ is $\sim 30$ eV (for
$\bar{Z}=4$) and the WDM behaves as a simple metal, with $\sigma$  dropping as
$T$ increases, and then increasing at higher $T_e$ when $\mu_e$ becomes
negative. The conductivity (for $T\leq 0.5 E_F$)  is determined mainly by the
value of $S(k)$ at $2 k_F$, shown in Fig.~\ref{carb-cond.fig}(b). This is set
by the C-C peak in $S(k)$, which  is relatively insensitive to $T$, and hence 
$\sigma$  is  also insensitive to temperature (compared to WDM Al or Li) in
this regime. The insensitivity of $S(k=2k_F$)to temperature  also leads to the strikingly
different  behavior of the ultrafast conductivity for  liquid carbon as
compared to $\sigma_{\rm uf}$ and $\sigma_{\rm ic}$ of  WDM-Al or Li. In
WDM-carbon the ultrafast and isochoric  conductivities are very close in
magnitude. The DFT+MD $\sigma_{\rm ic}$ values for 3.7 g/cm$^3$ differ
 from the NPA at
low-$T$ where strong-covalent bonds dominate. The $N\sim100$ atom DFT+MD simulations
may be seriously inadequate due to such C-C bond formation. The NPA itself
deals only in a spherically averaged way with the covalent bonding. That approximation
is probably sufficient for static conductivities if the bonding is truly transient.
In any case, accurate experimental $\sigma_{\rm ib}$ data for liquid carbon are badly needed.

\section{Conclusion}
Although it is not necessary in principle to distinguish between isochoric and
isobaric conductivities, as the specification of the density and temperature is
sufficient, the use of such a distinction is useful in comparing experiment
and theory. We see from our calculations that the temperature variations
of the three conductivities have distinct features.
Furthermore, the ultrafast conductivity is indeed a
physically distinct property as the ion subsystem remains unchanged
while only the electron subsystem is changed during the short time delay
between the pump pulse and the probe pulse. Thus in this study we have found it
useful to distinguish isochoric, isobaric and ultrafast  conductivities of
WDM systems, using Al, Li and C as examples.  The NPA $\sigma_{ib}$ are in
excellent agreement with the aluminum experimental data
 of Gathers~\cite{IJTher83},
while the DFT+MD+KG with 108-atom simulations estimate a lower conductivity. 
The NPA results are in moderate agreement with Oak Ridge $\sigma_{ib}$ for Li,
 as is also the case with DFT+MD+KG calculations. The carbon $\sigma_{ic},
\sigma_{uf}$ from NPA have a striking behaviour in the regime of (normal)
 densities studied here, and differ from Al and Li. We attribute this to
the effect of transient C-C bonds.

\appendix*

\section{}
$\,$\\
This appendix  addresses the following topics:\\
\begin{itemize}
\item  Neutral pseudoatom (NPA) calculation of the X-ray Thomson scattering (XRTS) 
ion feature $W(q)$ for comparison with the
density-functional-theory/molecular-dynamics (DFT+MD) calculations of Witte
{\it et al.}~\cite{Witte17}, where the excellent agreement is in clear contrast
to the disagreement for the conductivity datum for Li reported by Witte {\i et al.}
\item  Details of the neutral pseudoatom (NPA) model.
\item  Ziman formula for the conductivity using the NPA pseudopotential and the
ion-ion structure factor $S(k)$.
\item  Examples of DFT+MD and KG calculations for Al, Li, and C,
 and Drude fits to the KG conductivity of Al and Li.
\item  Review of the isobaric  and the isochoric conductivities of aluminum in the
 context of the experiment of Gathers, and the disagreement with the
 conductivity of Al reported in Fig.1 of Ref.~\cite{WittePRL17} by Witte {\it
 et al.} using the Heyd, Scuseria, and Ernzerhof (HSE) functional.
\end{itemize}
\begin{figure}[h]
\includegraphics[width=\columnwidth]{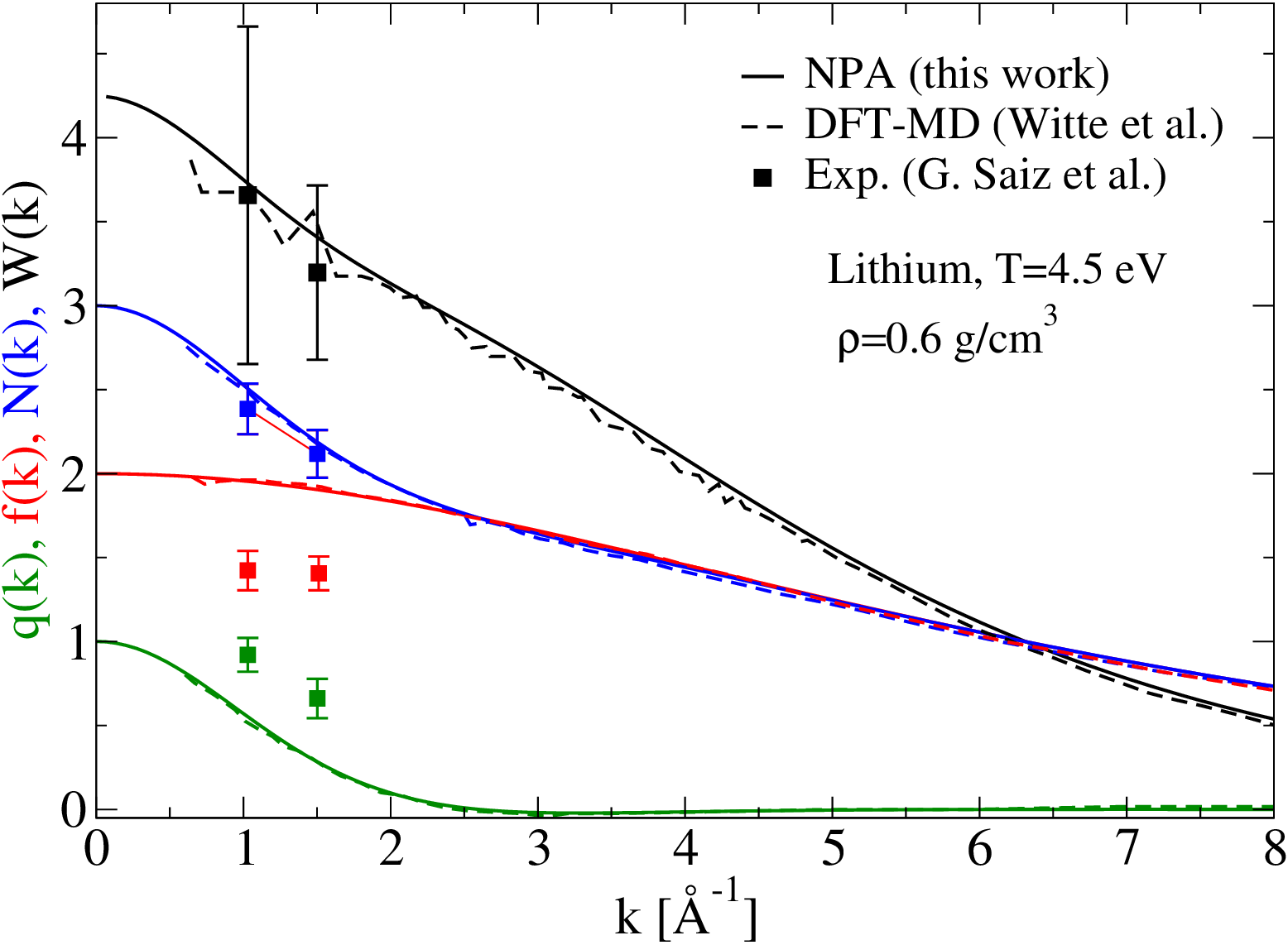}
\caption{(Color on line) Comparison of quantities relevant to XRTS calculated
using NPA+HNC (this work) and DFT+MD (Witte {\it et al.}~\cite{Witte17}) for
lithium. Values for $k$ smaller than about 0.6/\AA$\,$ are not available from
the DFT+MD simulation due to the finite size of the simulation cell. Here
$q(k)$ is the Fourier transform of the free-electron density at the Li ion in
the plasma, while $f(k)$ is the bound-electron form factor and
$N(k)=f(k)+q(k)$. The {\it ion feature} $W(k)=N(k)^2S(k)$ involves the ion-ion
structure factor $S(k)$. Experimental points are from Saiz {\it et al.} (2008)
cited in Witte {\it et al.}, Fig.\ 8}
\label{xrts.fig}
\end{figure}

\subsection{X-ray Thomson Scattering calculation for Li at
 density $\rho=0.6$ g/cm$^3$ and temperature $T=4.5$ eV.}

The calculation of XRTS of WDM using the NPA method has been
described in detail in Ref.~\cite{xrts-LH16}. The XRTS ion feature $W(k)$ for
Li at $T=4.5$ eV and $\rho=0.6$ g/cm$^3$ has been calculated (see
Fig.~\ref{xrts.fig})  to compare our NPA results with the results from the 
DFT+MD simulations by Witte {\it et al.} (Ref.~\cite{Witte17}, Fig.8). This
establishes the excellent agreement with the electronic structure part of the
NPA calculation and the ionic part,  $S_{ii}(k)$, resulting from the DFT+MD calculations,
{\it irrespective} of the exchange-correlation (XC) functional used. That is,
while we have used the local-density approximation (LDA)
 of the finite-$T$ XC functional $F_{xc}$ based on the
classical-map hyper-netted-chain scheme (CHNC)~\cite{PDWXC}, Witte {\it et al.}
have used the $T=0$ Perdew-Burke-Ernzerhof (PBE) XC functional~\cite{PBE96} which
includes gradient corrections. 

The mean ionization $\bar{Z}$ for Li obtained in the NPA is unity, in 
agreement with that used by Witte {\it et al.}. They calculate the quantities  
$q(k), f(k), N(k)$, and $W(k)=N(k)^2S(k)$. The quantity $q(k)$ is the
`screening cloud', i.e., the Fourier transform of the free-electron density at
the Li ion in the plasma, while $f(k)$ is the bound-electron form factor. Their
sum is denoted by $N(k)=f(k)+q(k)$. Finally, $W(k)=N(k)^2S(k)$ is the ion
feature and  involves the ion-ion structure factor $S(k)$.

The excellent accord between our XRTS calculation and that of Witte {\it et
al.} establishes that our $S(k)$, electron charge distributions, and potentials
$U_{ei}(k)$ and $V_{ii}(k)$ are fully consistent with the structure data and
electronic properties coming from DFT+MD. The $S(k)$ and $U_{ei}(k)$ are the only
inputs to the Ziman formula for $\sigma_0$.  Nevertheless, our estimate of the
conductivity disagrees strongly with the Kubo-Greenwood estimate of Witte {\it
et al.}$\,$. Given the relatively good agreement that we found with the
Oak Ridge experimental data, as well as with the Kietzmann data 
(see Fig.~\ref{Li-Kietz.fig}),
this disagreement is {\it a priori} quite surprising; one possible
contributory factor will be taken up in our discussion of the Kubo-Greenwood
formula, viz.,  that it may be caused by the use of a small 64-atom DFT+MD simulation
cell.

The conductivity estimate by Witte {\it et al.} for T=0.3 eV at 2.7 g/cm$^3$ is
also  problematic and it is taken up  below, in our discussion of Gathers' results
for aluminum.

\subsection{Details of the NPA model and $\bar{Z}$}

The NPA model used here~\cite{Pe-Be,eos95} has been described in many articles;
we summarize it again here for the convenience of the reader, as it should
not be assumed that it is equivalent to various  currently-available
ion-sphere (IS) average-atom (AA) models such as Purgatorio~\cite{SternZbar07} used in
many  laboratories. While these models are closely
related, they invoke additional considerations which are outside DFT.
We regard the NPA model as a rigorous  DFT model based on the variational
property of the grand potential $\Omega([n],[\rho])$ as a functional of both one-body
densities  $n(r)$ and $\rho(r)$, directly leading to two coupled KS equations where the
unknown quantities are the XC-functional for the electrons and the ion-correlation
functional for the ions~\cite{ilciacco93}. Approximations arise
in modeling those XC-functionals and decoupling the two KS equations for
simplified numerical work.

The NPA model assumes spherical
symmetry when dealing with fluid phases, and calculates the KS
states of a nucleus of charge $Z$ immersed in an electron gas of input density
$\bar{n}$. The ion distribution $\rho(r)$ is approximated by a neutralizing
uniform positive background containing a cavity of radius $r_{ws}$, with the
nucleus at the origin. The Wigner-Seitz (WS) radius $r_{ws}$ is  that of the
ion-density $\bar{\rho}$, i.e., $r_{ws}=\{3/(4\pi\bar{\rho)}\}^{1/3}$. The effect of the
cavity is subtracted from the final result where by the density response of a 
uniform electron gas to the nucleus is obtained. The validity
of this approach  has been
established in previous work, for the WDM systems investigated here and
 reviewed in Ref.~\cite{CPP}. The
solution of the KS equation extends up to $R_c=10 r_{ws}$, defining
a correlation sphere (CS) large enough for all electronic and  ionic
correlations with the central nucleus to have gone to zero. The WS cavity plays
the role of a nominal $\rho(r)$ to create a pseudoatom which is a neutral
scatterer and greatly facilitates the calculation.  The KS equations produce 
two groups of energy states, viz, negative and positive with respect to
the energy zero at $r\to \infty$ outside the CS. States in one group
 decay exponentially to
zero as $r\to R_c$, and in fact become negligible already for $r\to r_{ws}$ in
the case of low-$Z$ elements. These states, fully contained within the WS
sphere, are deemed bound states, and allow one to define a mean ionization per
ion, $Z_b=Z-n_b$, where $n_b$ is the total number of electrons in the bound
states and Z is the nuclear charge:
\begin{equation}
\label{nb-Z.eq}
Z_b=Z-n_b;\, n_b=\sum_{nl}2(2l+1)\int d\vec{r} f_{nl}|\phi_{nl}(r)|^2.
\end{equation}
Here $f_{nl}=1/\{1+\exp(x_{nl})\},\; x_{nl}=\{\epsilon_{nl}-\mu_0\}/T$
is the Fermi factor for the KS state $\phi_{nl}$ with energy
$\epsilon_{nl}$. The non-interacting electron chemical potential $\mu_0$ is
used here. Furthermore, there  are plane-wave-like phase-shifted KS states
which extend through the whole correlation sphere. These are continuum states
and their electron population is the free-electron distribution $n_f(r)$. The
nucleus $Z$, the bound electrons $n_b$, the cavity with a charge
$Z_c=(4\pi\bar{n}/3)r_{ws}^3$ and the free electrons form a neutral object and
hence it is a weak scatterer called the `neutral pseudoatom' (NPA). The Friedel sum
$Z_F$ of the phase shifts of the continuum states and the cavity charge $Z_c$
add up to zero when the KS-equations are solved self-consistently. Thus 
\begin{equation}
\label{Friedel.eq}
Z_c=Z_F=\frac{2}{\pi T}\int_0^\infty kf_{kl}k\{1-f_{kl}\}
 \sum_l(2l+1)\delta_l(k)dk.
\end{equation}
Here $f_{kl}$ is the Fermi occupation factor for the $k,l$-state with energy
$\epsilon=k^2/2$. Full self-consistency requires that
\begin{equation}
\label{self-cons.eq}
 Z_b=Z_c=Z_F,\;\; \bar{n}=\bar{Z}\bar{\rho}.
\end{equation}
Hence, given an input mean free-electron density $\bar{n}$, the WS radius
(equivalently $\bar{\rho}$) is iteratively adjusted till self-consistency is
obtained, i.e., Eq.~\ref{self-cons.eq} is satisfied to a chosen precision.  The
mean ionization  $\bar{\rho}$ is thus seen to be the Lagrange multiplier
ensuring charge neutrality, as first discussed in Ref.~\cite{DWP1}.  The
$\bar{\rho}$ resulting from the input $\bar{n}$ may not be the required
physical ion density, and hence several values of $\bar{n}$ and the
corresponding $\bar{\rho}$ are determined to obtain the actual $\bar{n}$ that
corresponds to the required experimental ion density $\bar{\rho}$. This process
produces a unique value of $\bar{Z}$, and the problem of having several
different estimates of $\bar{Z}$, as found in IS-AA
models~\cite{Murillo13,SternZbar07} does not arise here. The agreement among
$Z_F,Z_c, Z_b$ is essential to the convergence of the NPA-KS equations. It is
sensitive to the exchange-correlation (XC)-functional $F_{xc}(T)$ and to the
proper handling of self-interaction (SI) corrections, whenever $\bar{Z}$ is
close to a half-integer. Using a valid $\bar{Z}$ is essential to obtaining good
conductivities.

We emphasize that a key difference between IS models and the NPA is
that the free electrons are not confined to the Wigner-Seitz sphere, but move
in all of space as approximated by the correlation sphere. These differences are
discussed in Sec.~\ref{diffs-npa.sec}.

In this study we use the local-density approximation to the finite-$T$ 
XC-functional as parametrized by Perrot and Dharma-wardana~\cite{PDWXC}. This
simplest implementation (in LDA) is a useful reference step needed before more
elaborate implementations (involving SI, non-locality, etc.\  in the
XC-functionals) are used.

Since $\bar{Z}$ is the free-electron density per ion, it can develop
discontinuities whenever the ionization state of the element under study
changes due to, e.g., increase of $T$ or compression. This behaviour is
analogous to the formation or disappearance of band gaps in solids. In fact, if
the NPA model is treated with periodic boundary conditions, as for a solid with
one atom in the unit cell,  then the
discontinuity in $Z$ appears as the problem of correctly treating the formation
of a gap in the density of states (DOS) at the Fermi energy. A proper
evaluation of such features in the DOS and band gaps is difficult in DFT as 
this is a theory of the total energy as a functional of the one-body density,
{\it not} a theory of individual energy levels. The one-electron states are
given by the Dyson equation. Thus band-structure calculations inclusive of
GW-corrections are used in solids to obtain realistic band gaps and excitation
energies. In dealing with discontinuities in $\bar{Z}$, a similar procedure is
needed~\cite{PDW-levelwidth}, including the use of
self-interaction (SI) corrections and
 XC functionals that include  electron-ion correlation corrections, i.e., 
$F_{ei}(n,\rho)$~\cite{ilciacco93,Furutani90}.  

It should be mentioned that some authors have  claimed that $\bar{Z}$ ``does
not correspond to any well-defined observable in the sense of quantum 
mechanics''~\cite{PironBlenski11},i.e, that there is no quantum operator corresponding to
$\bar{Z}$. This view is incorrect as quantities like the
temperature $T$, the chemical potential $\mu$, and the mean ionization $\bar{Z}$
are quantities in quantum {\it statistical} physics. There may be no operator
for them in simple $T=0$ quantum theories. In most formulations of
quantum statistical physics these appear
as Lagrange multipliers related to the conservation of the energy, particle
number and charge neutrality. They can also be incorporated as operators in
more advanced field-theoretic formulations of statistical physics
(e.g., as in ``thermofield-dynamics'' of
Umezawa).  Some of these broader issues are discussed in Chapter 8 of
Ref.~\cite{apvmm13}.

Finally, it is noted that the mean number of electrons per ion, viz., $\bar{Z}$
in, e.g., gas-discharge plasmas, is routinely measured using Langmuir probes,
or derived from optical measurements of various properties including the conductivity
and the XRTS profile~\cite{GlenzerRedmer2009} for WDM-plasmas. Hence $\bar{Z}$
is a  well-established {\it measurable} property.

\subsection{Some Differences between the NPA model and typical
 average-atom models}
\label{diffs-npa.sec}
To our knowledge, no conductivity calculations 
using the
Purgatorio model  for {\it isobaric} aluminum are available
for comparison with 
experimental data. Such a comparison is also problematic due to the
lack of an unequivocal value for the mean ionization $\bar{Z}$
 in IS-AA models~\cite{SternZbar07}.
We list several differences with the NPA which particularly affect conductivity
calculations:

1. Most average-atom models are based on the IS-AA model where the
free-electron pileup around the nucleus is strictly
confined to the Wigner-Seitz sphere:
\begin{equation}
\label{mu0ws.eq} 
\bar{Z}=4\pi\int_0^{R_{ws}}\Delta n_f(r)r^2dr; \; \mbox{IS-AA model}.
\end{equation}
This condition, Eq.~\ref{mu0ws.eq}, was used in Salpeter's early IS model,
 in the Inferno model
of Lieberman, and in codes like Purgatorio~\cite{SternZbar07} derived from it,
 to determine
an  electron chemical potential $\mu_{\rm ws}^0$. It
is also used in  
Yuan {\it et al.}~\cite{Yuan96}, Faussurier {\it et al.}~\cite{Faussurier15},
Starrett and Saumon~\cite{StaSau13}, and in other AA codes discussed in
 Murillo {\it et al.}~\cite{Murillo13}.  However, $\mu_{\rm ws}^0$ is not identical
with the non-interacting $\mu_0$ because it includes a confining potential applied
to the free electron density $n_f(r)$ constraining the electrons to the IS.
 As it is applied via a boundary condition, 
it is a non-local potential. The KS XC potential is also a non-local potential
 and hence the use of Eq.~\ref{mu0ws.eq} contaminates the XC potential. 
On the other hand, DFT is based on mapping the
interacting electrons to a system of {\it non-interacting electrons} whose chemical
potentially is rigorously $\mu_0$, as used in the NPA model that we employ. In the NPA
we use a CS with a large radius $R_c$.
\begin{equation}
\label{mu0NPA.eq} 
\bar{Z}=\pi\int_0^{R_{c}}\Delta n_f(r)r^2dr; \; \mbox{NPA model}.
\end{equation}
The upper limit of the integral is $R_c\approx 10 r_{ws}$ and hence deals with a
sphere  large enough for all correlations with the central ion to have died down at the surface of the sphere.
This enables the use of the non-interacting chemical potential in the NPA, as needed
in DFT, since all equations use the large-$r$ limit beyond the CS as the reference state.

The constraint placed by Eq.~\ref{mu0ws.eq} is clearly invalid at low temperatures
where the de Broglie wavelength of the electrons, being proportional to $1/\surd{T}$,
exceeds $r_{ws}$ at sufficiently low $T$. Hence such AA-models become invalid
 at low temperatures and are not true DFT models. In contrast, the first
 successful applications
of the NPA (in the 1970s) were  to low-temperature solids.

2. The use of the constraint placed by Eq.~\ref{mu0ws.eq} in AA models
has far reaching consequences
as it prevents the possibility of providing a unique definition of the mean ionization,
as emphasized by Stern {\it et al.}~\cite{SternZbar07} in regard to the Inferno code.
In fact even at high temperatures, there are at least two definitions of $\bar{Z}$ that
differ, and hence the estimates of the electrical conductivity are not unambiguous. This
is not the case in the NPA. The problem of discontinuities in $\bar{Z}$ and the
under-estimate of bandgaps by DFT theory were already discussed in the previous subsection.

3. The IS-AA models do not satisfy a Friedel sum rule for $\bar{Z}$, while the
 $f$-sumrule is also constrained by the condition imposed by Eq.~\ref{mu0ws.eq}.

4. As the electrons are confined to the WS sphere in  IS-AA models, they cannot
display pre-peaks due to transient covalent bonding as found in 
liquid carbon, hydrogen and other low-$Z$ WDMs. This was confirmed by
 Starrett {\it et al.}~\cite{StaSauDaligHam14} for carbon
for their AA model. The bonding
occurs by an enhanced electron density in the inter-ionic region between two WS spheres,
and this is not allowed in IS models.
In contrast, the NPA model shows pre-peaks in $g_{ii}(r)$ corresponding to transient
C-C bonding in liquid carbon, and produces a pair-potential with a minimum corresponding
to the C-C covalent bond distance at sufficiently low $T$~\cite{cdw-carbon16}. Similar
pre-peaks are found via NPA calculations for  warm-dense hydrogen and low-$Z$ elements
 in the appropriate  temperature and density regimes~\cite{HugCHNC17}.

\subsection{Pseudopotentials and pair-potentials from the NPA}
\label{pots-npa.sec}
The KS calculation for the electron states for the NPA in a fluid involves
solving a simple radial equation. The continuum states $\phi_{k,l}(r),
\epsilon_k=k^2/2$, with occupation numbers $f_{kl}$, are evaluated to a
sufficiently large energy cutoff and for an appropriate number of $l$-states
(typically 9 to 39 were found sufficient for the calculations presented here).
 The very high-$k$ contributions are included
by a Thomas-Fermi correction. This leads to an evaluation of the
free-electron density $n_f(r)$, and the free-electron density pileup $\Delta
n'(r)=n_f(r)-\bar{n}$. A part of this pileup is due to the presence of the
cavity potential. This contribution $m(r)$ is evaluated using its linear
response to the electron gas of density $\bar{n}$ using the interacting
electron response $\chi(q,T_e)$. The cavity corrected free-electron pileup
$\Delta n_f(r)=\Delta n'(r)-m(r)$ is used in constructing the electron-ion
pseudopotential as well as the ion-ion pair potential $V_{ii}(r)$ according to
the following equations (in Hartree atomic units) given  for
Fourier-transformed quantities:
\begin{eqnarray}
\label{pseudo.eq}
U_{ei}(k) &=& \Delta n_f(k)/\chi(k,T_e),\\
\label{resp.eq}
\chi(k,T_e)&=&\frac{\chi_0(k,T_e)}{1-V_k(1-G_k)\chi_0(k,T_e)},\\
\label{lfc.eq}
G_k &=& (1-\kappa_0/\kappa)(k/k_\text{TF});\quad V_k =4\pi/k^2,\\
\label{ktf.eq}
k_{\text{TF}}&=&\{4/(\pi \alpha r_s)\}^{1/2};\quad \alpha=(4/9\pi)^{1/3},\\
\label{vii.eq}
 V_{ii}(k) &=& Z^2V_k + |U_{ei}(k)|^2\chi_{ee}(k,T_e).
\end{eqnarray}
Here $\chi_0$ is the finite-$T$ Lindhard function, $V_k$ is the bare Coulomb
potential, and $G_k$ is a local-field correction (LFC). The finite-$T$
compressibility sum rule for electrons is satisfied since $\kappa_0$ and
$\kappa$ are the non-interacting and interacting electron compressibilities
respectively, with  $\kappa$ matched to the $F_{xc}(T)$ used in the KS
calculation. In Eq.~\ref{ktf.eq}, $k_\text{TF}$ appearing in the LFC is the
Thomas-Fermi wavevector. We use a $G_k$ evaluated at $k\to 0$ for all $k$
instead of the more general $k$-dependent form (e.g., Eq.~50  in
Ref.~\cite{PDWXC}) since the $k$-dispersion in $G_k$ has negligible effect for
the WDMs of this study. Steps towards a theory using self-interactions
corrections in the $F_{xc}$, a modified electron DOS, self-energy corrections
etc., have also been given~\cite{PDW-levelwidth}. In this study  we use the above
equations, and only in the LDA.

\subsection{Calculation of the ion-ion Structure factor}

The ion-ion structure factor $S(k)$ is also a first-principles quantity as it
is calculated using the ion-ion pair potential, Eq.~\ref{vii.eq} given above.
For simple fluids like aluminum we use the modified hyper-netted-chain (MHNC) 
equation. 
\begin{eqnarray}
\label{MHNC.eq}
g(r)&=&\exp\{-\beta V_{ii}(r)+ h(r)-c(r)+ B(r)\}, \\
\label{OZ.eqn}
h(r)&=& c(r)+\bar\rho\int d\vec{r}_1 h(\vec{r}-\vec{r}_1)c(\vec{r}_1), \\
h(r)&=& g(r)-1.
\end{eqnarray}
Here $c(r)$ is the direct correlation function. Thermodynamic consistency
(e.g., the virial pressure being equal to the thermodynamic pressure) is
obtained by using the Lado-Foiles-Ashcroft (LFA) criterion (based on the
Gibbs-Bogoliubov bound for the free energy) for determining $B(r)$ using the
hard-sphere model bridge function~\cite{LFA83}. That is, the hard-sphere
packing fraction $\eta$ is selected according to an energy minimization that
satisfies the LFA criterion. The iterative solution of the MHNC equation, i.e.,
Eq.~(\ref{MHNC.eq}), and the Ornstein-Zernike (OZ) equation,
Eq.~(\ref{OZ.eqn}), yield a $g_{ii}(r)$ for the ion subsystem. The LFA
criterion and the associated hard-sphere approximation can be avoided if
desired, by using MD with the  pair potential to generate the $g(r)$. The
hard-sphere packing fraction $\eta$ calculated via the LFA criterion is the
only parameter extraneous to the KS scheme used in our theory. In
calculating the $S(k)$ of complex fluids like carbon, where the leading peak in
$g(r)$ is {\it not} determined by packing effects but by transient C-C bonding,
we  use the simple HNC equation.

\subsection{Calculation of the electrical conductivity}
The electrical conductivity is calculated from the numerically convenient form
of the Ziman formula  given in Ref.~\cite{eos95}. The Ziman formula is
sometimes derived from the Boltzmann equation. However, the KG formula and also
the Ziman formula can both be derived from the Fermi golden
rule~\cite{res2006}. The Ziman formula uses the `momentum-relaxation time'
approximation, while the KG formula typically uses the same approximation when
extracting the static conductivity using a Drude fit to the dynamic
conductivity $\sigma(\omega)$. The Ziman formula used here is:
\begin{eqnarray}
\sigma&=&1/R\; R=(\hbar/e^2)(3\pi\bar{n}\bar{Z})^{-1}I\\
I&=&\int_0^\infty \frac{q^3\Sigma(q)dq}{1+\exp\{\beta(\epsilon_q/4-\mu)\}},\\
\epsilon_q &=&(\hbar q^2/2m),\;\beta=1/T,\\
\Sigma(a)&=&S(q)|U_{ei}(q)/\{2\pi\varepsilon(q)\}|^2,\\
1/\varepsilon(q&)=&1+V_q\chi(q,T).
\end{eqnarray}
The ``Born-approximation-like'' form used here is valid to the same extent that
the pseudopotential $U_{ei}(q)$ constructed from the (non-linear) KS
$n_f(r)$ via linear response theory  (Eq.~\ref{resp.eq}) is valid. The $S(k)$
used are  available even for small-$k$ unlike in DFT+MD simulations where the
smallest accessible $k$-value is limited by the finite size $L_{bx}$ of the
simulation cell.
\begin{figure}[t]
\includegraphics[width=\columnwidth]{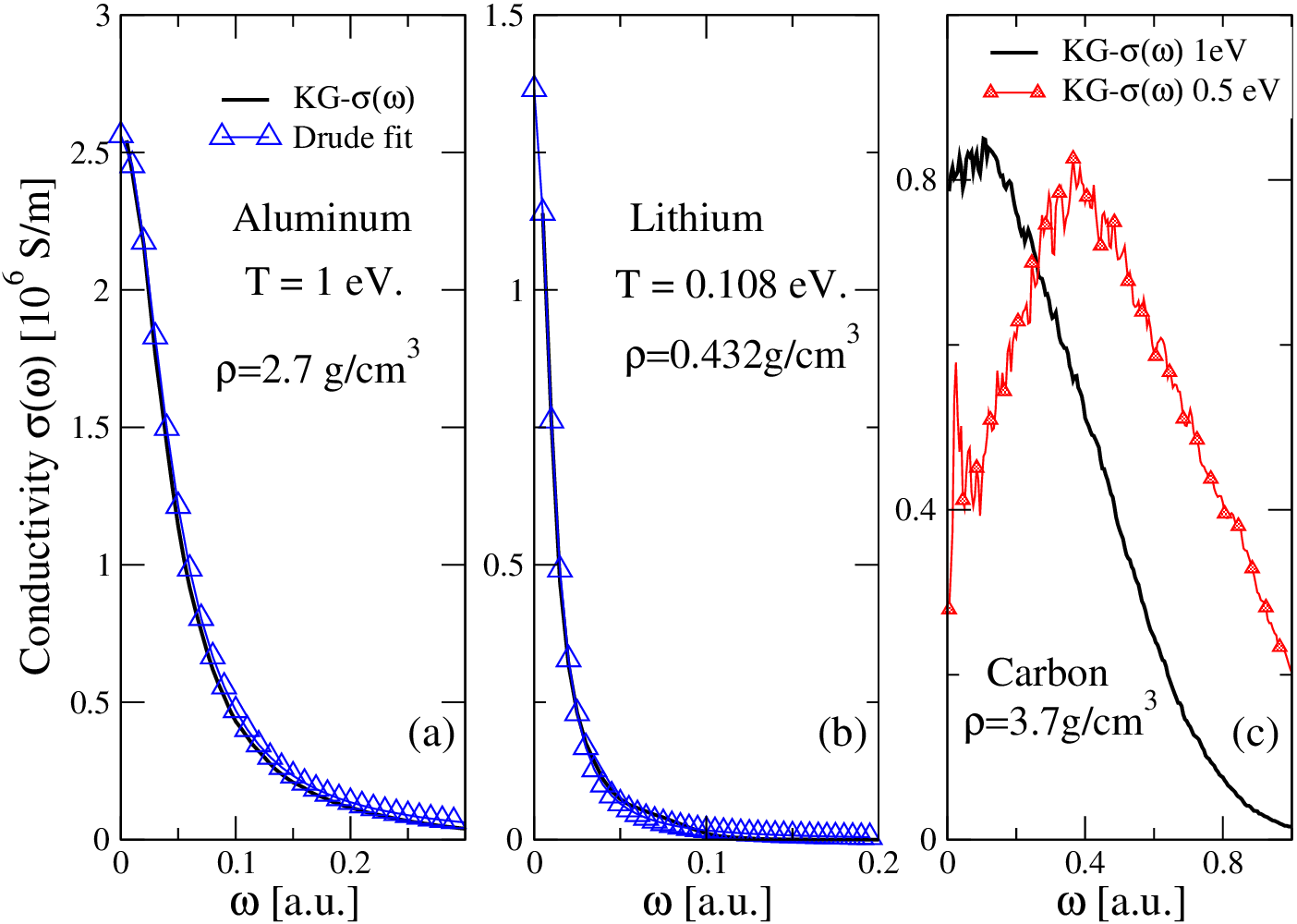}
\caption{(Color online) KG conductivity $\sigma(\omega)$ for Al, Li, and C.
Note the slight non-Drude behaviour of Li $\sigma(\omega)$ near 0.08 a.u. in
panel (b). The carbon $\sigma(\omega)$ is highly non Drude-like, with the peak
moving to higher energy as $T$ is lowered; no Drude form is shown for carbon.}
\label{drude.fig}
\end{figure}

\subsection{The Kubo-Greenwood conductivity}
\label{KG.sec}
The KG dynamic conductivity $\sigma(\omega)$ is a popular approach to
determining the static conductivity of WDM systems via
DFT+MD~\cite{Recoules02}. In our simulations we have used  $N$=108 atoms in the
simulation cell, with a $2\times2\times2$ Monkhorst-pack $k$ grid; the PBE XC
functional was used. The energy cutoff was taken to be sufficiently high
that the occupations in the highest KS states were virtually negligible.
The quenched-crystal KS-eigenstates $\phi_\nu(r)$ and eigenvalues
$\epsilon_\nu$, where $\nu$ is a band-index quantum number, are used in the
Kubo-Greenwood conductivity as provided in the standard ABINIT code. Usually six
to ten such evaluations were obtained by evolving the quenched crystal by
further MD simulations (using only the $\Gamma$ point), and in each case the
$\sigma(\omega)$ was obtained -- see Fig.~\ref{drude.fig} for typical
aluminum, lithium and carbon results for $\sigma(\omega)$.

The aluminum $\sigma(\omega)$ is well-fitted by the Drude form: 
\begin{equation}
\sigma(\omega)=\sigma_0/(1+(\omega\tau)^2), \sigma_0=\bar{n}\tau.
\end{equation}
However, there is no justification for using a Drude form for carbon. The peak
position in $\sigma(\omega)$ roughly corresponds to the `bonding $\to$
antibonding' transition in the fluid containing significant covalent bonding
(see Fig.\ 4(b) of the main text) at 0.5 eV. This is seen from the strong peak
in $g(r)$ near 3 a.u. (1.55 \AA$\,$) corresponding to the C-C bond length. 
This suggests that the $N=108$ simulation is quite inadequate for complex
liquids like carbon, as bonding reduces the effective $N$ of the simulation. In
the case of carbon, the static limit of the KG $\sigma(\omega)$ was simply
estimated from the trend in the $\omega \to 0$ region rather than using a Drude
fit. Furthermore, the different quenched crystals (108 atoms in the
simulation) gave significant statistical variations, as reflected in the error
bars shown in Fig.\ 4(a) of the main text. At higher $T$, e.g., for $T=1-2$ eV,
the estimated conductivity behaves similar to that from the NPA, but somewhat
less conductive. The KG formula does not include any self-energy corrections in
the one-electron states and excitation energies, and less importantly, no
ion-dynamical contributions either, as the ions are stationary
(Born-Oppenheimer approximation). The form of $\sigma(\omega)$ including ion
dynamics has been discussed by Dharma-wardana at the Carg\`{e}se NATO work shop
in 1992~\cite{carges}. 

\subsection{The  conductivity of Li at $T=$4.5 eV and density 0.6 g/cm$^3$}
The conductivity of Li, at density $\rho=$ 0.6 g/cm$^3$ at 4.5 eV estimated by Witte {\it
et al.}~\cite{Witte17}, is roughly a factor of five less than that obtained from
NPA+Ziman. While the NPA calculation may differ from another calculation
by, at worst, a factor of 2, it is hard to find an explanation for this
strong disaccord, given the good agreement in the XRTS calculation.
 One possibility is the use of a 64-atom cell in DFT+MD 
 for Li at a chemical potential $\mu\sim 0$.
DFT+MD and KG using $N\sim 100$ atoms in the simulation seems to significantly
underestimate $\sigma_0$ for low-valence substances like Li, Na, especially as
$T$ is increased. Low-valence materials have a small $\mu=E_F$ and hence a
modest increase in $T$ can push $\mu$ to small values where small-$k$
scattering is important, and finally to  $\mu<0$  values (classical regime).

At low $T/E_F$ the major contributions to $\sigma$ are provided by electron
scattering between $-k_F$ and $k_F$, $k_F=\sqrt{2E_F}$, i.e., momentum changes
of the order of 2$k_F$. However, at finite $T$, $\mu$ replaces $E_F$, and as
$T$ increases $\mu \to 0$ and to negative values. The scattering momenta near
$\mu\to +0$ are in the small-$k$ region, These contribute significantly to
$\sigma$ at $T=4.5$ eV for Li at 0.6 g/cm$^3$. In Li, if a 64-atom simulation
is used, an appropriate length $a$ of the cubic simulation cell would be
$a=20.26$ a.u. The smallest momentum accessible by such a simulation is
$\pi/a$=0.16/(a.u.) and hence the corresponding Kubo-Greenwood formula will not
sample the small $k<0.16$ region.  We see from Fig.~\ref{xrts.fig}  also that
the DFT+MD simulations do not provide values for $k$ smaller than
$\approx0.6$/\AA$\,$  due to the finite size of the cell  used in
Ref.~\cite{WittePRL17}.

Hence such DFT+MD+KG calculations of $\sigma$ are strongly weighted to the
larger-$k$ strong scattering regime and predict a low conductivity. 
 The results of Pozzo {\it et al.}, where a 1000-atom simulation was
 needed for Na is a case in point. However,
such large simulations are beyond the scope of many
 laboratories while NPA-type approaches
usually provide results to within a factor of two in the worst case.

\begin{figure}[t]
\includegraphics[width=\columnwidth]{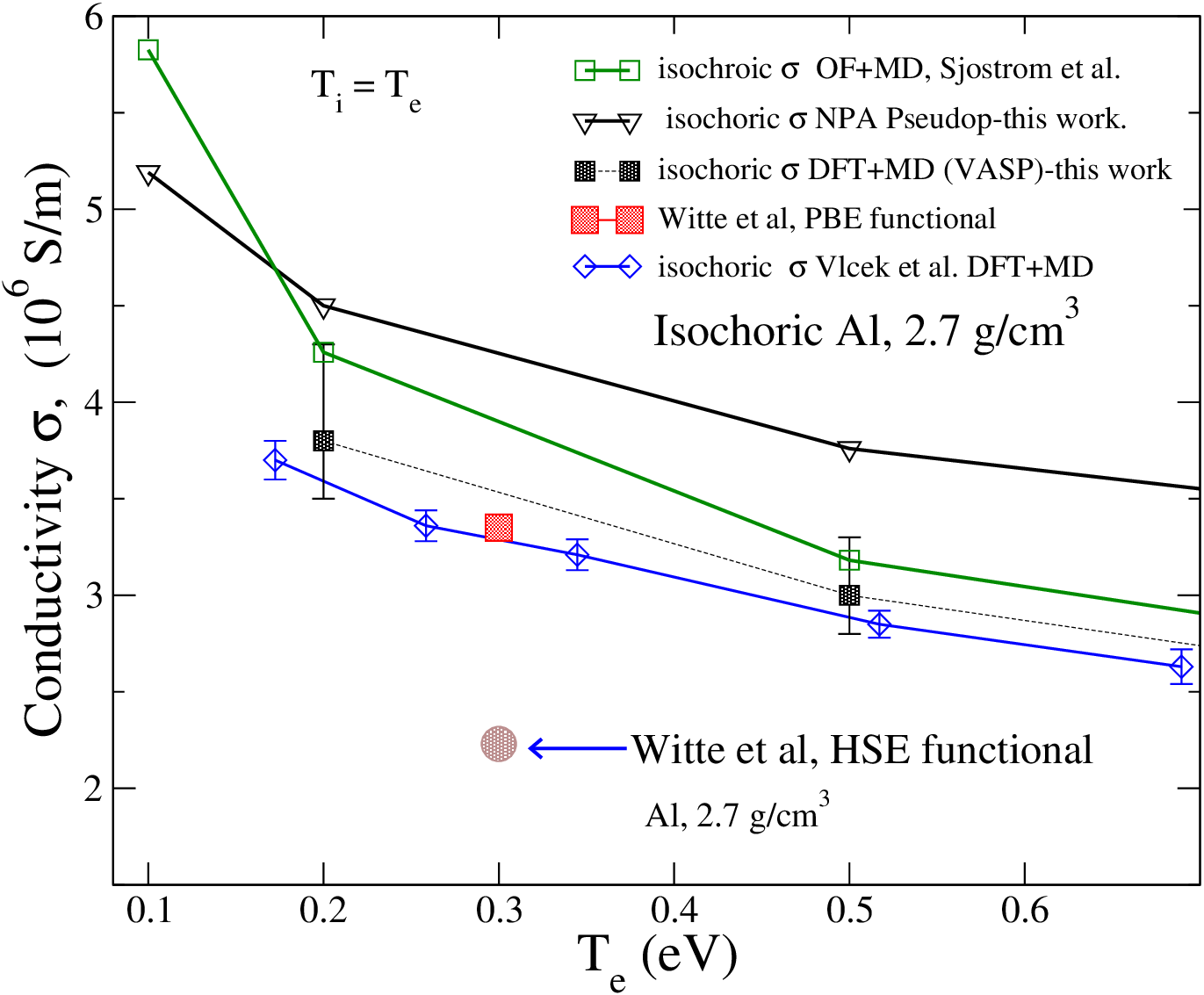}
\caption{(Color online) Isochoric conductivity of aluminum from near its
melting point to about 0.7 eV, expanded from Fig.\ 1 of the main text, and now
including the Witte {\it et al.}~\cite{WittePRL17} calculation of the
Al-conductivity at 0.3 eV and $\rho=2.7$ g/cm$^3$. Our DFT+MD data and those 
of Vl\v{c}ek are shown.
} 
\label{al-sig-isoch.fig}
\end{figure}

\subsection{Isobaric and isochoric conductivity of aluminum
 in the liquid-metal region}
High-quality  experimental data (errors of $\pm$ 6 \%) are available for the
isobaric conductivity $\sigma_{\rm ib}$ of liquid aluminum at low
$T$~\cite{IJTher83, Gathers86}. The relevant  region, viz.,
 (a) of Fig.~1 of the main text, is shown
enlarged to display the experimental and calculated data in
 Fig.~\ref{al-sig-isob.fig}.
 The NPA calculation
is in excellent agreement with the experiment of Gathers, to  well within the
error bars. On the other hand, the DFT+MD calculation captures about 75\% of
the experimental conductivity. A $\sim$100-atom simulation cannot capture the
$k$-values smaller than $\pi/a\sim 0.12$/(a.u.) for Al at this density,  and
may contribute to some of the under-estimate. 

Isochoric conductivities (with $\rho=2.7$ g/cm$^3$) of aluminum obtained from
the NPA and from DFT+MD by us and by Vl\v{c}ek {\it et al.}~\cite{Vlcek12} are
shown in Fig.~\ref{al-sig-isoch.fig},  together with a single data point
 from Witte {\it et al.}~\cite{WittePRL17} with the PBE functional, and with
 the HSE functional. The result obtained
 using the HSE XC-functional  is a strong
 underestimate compared to
other DFT+MD~\cite{DejaKreCol02,Recou05}, the orbital-free calculation
 and the NPA estimates.

In Ref.~\cite{WittePRL17} Witte {\it et al.} 
strongly argue for the HSE functional even for aluminum, a `simple' metal proven to
work well with more standard approaches. The value of 2.23 $\times 10^6$ S/m
quoted by them at 0.3 eV, 2.7 g/cm$^3$, is taken to agree with experiment,
based on their interpretation of the experimental data of
Gathers~\cite{IJTher83}. However, as discussed below, 
Gathers' datum at 0.3 eV ($\simeq$3500K) is  for
isobaric aluminum  at $\rho=1.875$ g/cm$^3$ and 0.3 GPa.

\subsection{The experimental data of Gathers}
Gathers measures the resistivity of aluminum in an isobaric experiment,
starting from the solid ($\rho_0$=2.7 g/cm$^3$, $v_0$=0.37 cm$^3$/g) and
heating to the  range 933K to 4000K at 0.3 GPa~\cite{IJTher83}.  Gathers himself
recommends the Gol'tsova-Wilson~\cite{Goltsova65, Wilson69} volume 
expansion data rather than those measured by him. 
 In Table II of the 1983 publication of
Gathers~\cite{IJTher83}, the experimental resistivity (``raw
 data'') calculated
using the nominal enthalpy input to the sample is given in column 5. The
apparatus and the sample undergo volume expansion; the resistivities
for the input enthalpy corresponding to the volume
expanded sample (using the Gol'tsova-Wilson data) are given in column 4 of
 the same table. Hence the ``volume corrected''
isobaric resistivity for aluminum in the range ($T$=993K, $\rho$=2.42g/cm$^3$)
to ($T$=4000K, $\rho$=1.77g/cm$^3$) are the values found in column 4 while
 column 5 gives  the ``raw data''. Column 4 resistivities agree with the
 isobaric resistivity values that may also be obtained from the fit
 formula given in the last row of Table 23 of the 1986 Gathers review\cite{Gathers86}.

Since Table II as given by Gathers is somewhat misleading, we have recalculated
the resistivities $R$ using the fit equations given by Gathers. Eq. (8) gives
the (expansion-uncorrected) ``raw data'', labeled $R_\text{G}$. 
 The expansion correction essentially
brings  the input heat to the actual volume of the sample. Thus equation (9),
 where the enthalpy input is corrected for volume  expansion
 agrees with Gathers'
fit equation given subsequently in 1986~\cite{Gathers86} and
 hence labeled $R_{1986}$.
Gathers uses the enthalpy as the primary variable in equation (8) and (9),
 but also gives $R_\text{G}$ directly as a function of $v/v_0$ in equation (10).
 Thus Eqs.~(8) and (10) yield the same resistivity $R_\text{G}$ at a given
density and corresponding $T$, while Eq. (9) is the volume-corrected equation 
restated  in the 1986 review.

\begin{table}
\caption{\label{revisedgathers.tab}
Gathers' data for Al
recalculated from his 1983 fit equations (6)-(10)
and also from his fit equation (reproduced as Eq.~\ref{gathers86.eq})
 given  in the last row of Table 23 of the 1986 review~\cite{Gathers86}}
\begin{ruledtabular}
\begin{tabular}{ccccccc}
T   & $v/v_0$ & $\rho$& $R_\text{G}$ & $R_{\rm Eq.(9)}=R_{1986}$ & $\sigma_{\rm ib}$ \\

(K) &  ---    & (g/cm$^3$) & $(\mu\Omega$m) & ($\mu\Omega$m) & $10^6$ (S/m) \\
  &  Eq.(7)   &   & Eq.(8) & Eq.(9) & $1/R_{\rm Eq.(9)}$  \\

Gathers & Col.3& ---  & Col.5  & Col.4  &$1/R_{1986}$  \\
\hline \\
933  &  1.12  & 2.42  & 0.261  & 0.233  & 4.30 \\
1000 &  1.12  & 2.41  & 0.268  & 0.238  & 4.20 \\
1500 &  1.18  & 2.29  & 0.331  & 0.281  & 3.56 \\
2000 &  1.24  & 2.18  & 0.370  & 0.324  & 3.08 \\
2500 &  1.30  & 2.08  & 0.476  & 0.367  & 2.73 \\
3000 &  1.37  & 1.97  & 0.560  & 0.409  & 2.44 \\
3500 &  1.44  & 1.87  & 0.651  & 0.451  & 2.22 \\
4000 &  1.52  & 1.77  & 0.751  & 0.494  & 2.03 \\
\end{tabular}
\end{ruledtabular}
\end{table}

According to Gathers, the experimental resistivities have an
 error of $\sim\pm 6$\%.
The relevant  equations from Gathers' 1983 work are given below:
\begin{eqnarray}
H&=& 0.0048910+0.0010704 T\\
 & &+2.3084\cdot 10^{-8}T^2 ,\nonumber {\rm Gathers~Eq.} (6),\\
v/v_0&=&1.0205+8.3779 \cdot 10^{-2}H\\
     & &+4.9050 \cdot 10^{-3} H^2,\nonumber {\rm Gathers~Eq.} (7),\\
R_{9}&=&0.1494+7.9448\cdot 10^{-2} H\\
 & &-1.3189\cdot 10^{-3}  H^2,\nonumber
 {\rm Gathers~Eq.} (9),\\ 
1.12 &\le& v/v_0\le 1.56 ; \;\;\; {\rm i.e.}\\
     &  & 2.411 \,{\rm g/cm}^3\le {\rm density}\le 1.731 \,{\rm g/cm}^3.\nonumber 
\end{eqnarray}
The enthalpy $H$ can be eliminated in Gathers' Eq.~(9), i.e. (A.22),
 using the preceding equations. The result agrees with
the fit equation given in the subsequent 1986 review article~\cite{Gathers86},
Table 23 (last row). This is given as a fit for the isobaric
 resistivity (at 0.3 GPa) 
($\mu\Omega$ m), viz.,
\begin{equation}
\label{gathers86.eq}
R(v)=-1.0742+4.1997\times10^3\cdot v-2.5124\times 10^6\cdot v^2. 
\end{equation}
Here $v$ is the volume in  m$^3$kg$^{-1}$ with
 4.1$\times 10^{-5}\le v\le 5.78\times^{-4}$.
The resistivity calculated from this equation agrees  with column
4 of Table II of Gathers~\cite{IJTher83}.


The NPA calculation which takes the nuclear charge,
 temperature and density as
the only inputs and uses the finite-$T$ PDW XC-functional(LDA)~\cite{PDWXC}
 gives excellent
agreement for $\sigma_{\rm ib}$ with the Gathers' data at all densities listed
 in Table~\ref{revisedgathers.tab}, as seen in
Fig.~\ref{al-sig-isob.fig}.
At  $T=0.3$ eV, $\rho=$1.875 g/cm$^3$ $\sigma_{\rm ib}=2.22 \times 10^6$ S/m,
while the HSE functional used with MD+DFT+KG gives this conductivity only
at 2.7 g/cm$^3$, as reported  by Witte {\it et al.}~\cite{WittePRL17}.

Our DFT+MD estimates of the isochoric conductivity using the PBE functional, 
the DFT+MD estimates of Vl\v{c}ek {\it et al.}, and the Witte {\it et al.}
DFT+MD estimate~\cite{WittePRL17} using the PBE functional for 2.7 g/cm$^3$ at
0.3 eV are in close agreement. They all fall below the NPA+Ziman estimate, and
we attribute this partly to the inability of the DFT+MD+KG approach to access
small-$k$ scattering contributions unless the number of atoms $N$ in the
simulation  is sufficiently large. Furthermore, as $T/E_F\to 0$, the estimate
of the derivative of the Fermi function and also the matrix-element of the
velocity operator probably require an increasingly more dense mesh of $k$-points.

\end{document}